\documentclass[%
reprint,
nofootinbib,
 amsmath,amssymb,
 aps,
]{revtex4-2}
\usepackage{stmaryrd}
\usepackage{empheq}
\usepackage{graphicx}
\usepackage{dcolumn}
\usepackage{bm}
\usepackage{hyperref}

\usepackage[caption=false]{subfig}
\usepackage{graphicx}
\begin{document}
\preprint{APS/123-QED}

\title{Searching for a waveform-agnostic gravitational wave signal in pulsar timing arrays}
\author{Heling Deng}
\affiliation{Columbia Astrophysics Laboratory, Columbia University, New York, NY 10027, USA}

\author{Bence Bécsy}
\affiliation{Institute for Gravitational Wave Astronomy and School of Physics and Astronomy, University of Birmingham, Edgbaston, Birmingham B15 2TT, UK}

\author{Yuri Levin}
\affiliation{Physics Department and Columbia Astrophysics Laboratory, Columbia University, New York, NY 10027, USA}
\affiliation{Department of Physics and Astronomy, Monash University, Clayton, VIC 3800, Australia}

\author{Neil J. Cornish}
\affiliation{eXtreme Gravity Institute, Department of Physics, Montana State University, Bozeman, MT 59717, USA}

\author{Xavier Siemens}
\affiliation{Department of Physics, Oregon State University, Corvallis, OR 97331, USA}

\begin{abstract}

Pulsar timing arrays have recently provided compelling evidence for a nanohertz stochastic gravitational wave background, motivating searches for gravitational waves from localized sources. Most existing searches assume specific waveform templates, which can be computationally demanding and potentially insensitive to unexpected signals. We introduce a waveform-agnostic framework that models signal-induced timing residuals via a Fourier expansion. A Lorentzian hyperprior is imposed on the variances of the Fourier coefficients, providing a flexible spectral envelope that captures the signal’s dominant frequency and bandwidth while remaining agnostic to its exact shape. Analytical marginalization over the Fourier coefficients then yields a Bayesian hierarchical framework that concurrently infers the source sky location, its frequency content, and the stochastic background. To mitigate contamination from unmodeled pulsar noise, we further allow for additional flat-spectrum features for each pulsar. Tests on simulated datasets show that the method is robust and provides a flexible tool for future PTA searches, with sensitivity to both expected and unexpected gravitational wave phenomena.

\end{abstract}
\maketitle

\section{Introduction}
Recent results from pulsar timing array (PTA) collaborations have provided strong evidence for a stochastic gravitational wave background (SGWB) in the nanohertz band \cite{NANOGrav:2023gor,EPTA:2023fyk,Reardon:2023gzh,Xu:2023wog,miles2025meerkat}. This background may arise from the superposition of signals from a population of supermassive black hole binaries (SMBHBs), or from processes in the early universe \cite{NANOGrav:2023hvm,Ellis:2023oxs,Figueroa:2023zhu}. Beyond the stochastic background, PTAs are also sensitive to strong individual sources. Among the most studied targets are continuous waves from nearby SMBHBs \cite{Sesana:2008xk,Rosado:2015epa,Kelley:2017vox,Becsy:2022pnr,Yardley:2010kv,NANOGrav:2014zwv,Zhu:2014rta,Babak:2015lua,Aggarwal:2018mgp,NANOGrav:2023bts,IPTA:2023ero,Gundersen:2024qmq}, but a wide range of other phenomena, such as bursts from cosmic string cusps \cite{Damour:2000wa,Damour:2001bk,Yonemaru:2020bmr,xia2025searching}, highly eccentric black hole encounters \cite{Finn:2010ph, Dandapat:2024ipc}, and the permanent spacetime displacement known as gravitational wave memory \cite{Pshirkov:2009ak,vanHaasteren:2009fy,Seto:2009nv,Cordes:2012zz,Madison:2014vca,Wang:2014zls,NANOGrav:2015xuc,NANOGrav:2019vto,Sun:2022hvp,agazie2025nanograv,Tomson:2025oox,tomson2026search}, may also produce detectable signatures.

Conventional searches for individual sources typically rely on specific waveform templates. While statistically robust, these analyses can be computationally demanding (due to a large parameter space or complex waveforms) and insensitive to unexpected or exotic signals. This has motivated the development of ``waveform-agnostic'' searches designed to identify coherent timing residuals across the array without prior knowledge of the signal's shape.

Several waveform-agnostic frameworks have been developed over the years, particularly focusing on generic burst signals in PTA data. In Ref. \cite{Finn:2010ph}, an analytical hybrid frequentist-Bayesian approach was introduced, where each data point was assigned a parameter describing the burst shape, and a maximum a posteriori estimate was used to fix the hyperparameters. This approach was later extended in Ref. \cite{Deng:2014kua} through a Bayesian nonparametric analysis. In Refs. \cite{Zhu:2014rta,Zhu:2015tua,Madison:2015txa}, frequentist frameworks were developed in both the time and frequency domains, where piecewise linear functions were used to describe the signal, and a least-squares fitting process was performed to estimate the waveform parameters. More recently, some authors of the present paper introduced a Bayesian framework in which the signal is modeled by linear interpolation, with the interpolation coefficients analytically marginalized in the likelihood \cite{Deng:2023btv}. This formulation greatly reduces the effective dimensionality of the parameter space, leading to computationally efficient analyses. Another sophisticated Bayesian approach was presented in Refs. \cite{Becsy:2020utk,Taylor:2024cgu}, where both the generic burst signal and unmodeled pulsar noise are modeled as superpositions of Morlet-Gabor wavelets, allowing for flexible modeling of localized time-frequency structures.

These previous studies have largely focused on capturing generic signal features without explicitly incorporating the underlying physical characteristics of the source.\footnote{Possible exceptions include Refs. \cite{Becsy:2020utk,Taylor:2024cgu}, which use sine-Gaussian wavelets to capture short-duration burst signals.} In this work, following a similar philosophy to Ref. \cite{Deng:2023btv}, we introduce a waveform-agnostic framework based on linear regression. Compared with previous studies, our Bayesian hierarchical modeling is both computationally efficient and capable of capturing the frequency content of the potential signal. 

We model the signal-induced timing residuals using a Fourier expansion at discrete frequencies $k/3T$, where $k=1,2,\ldots$, and $T$ is the PTA observation span. The extended period $3T$ allows for the modeling of non-periodic signals. The frequency content of the signal is encoded in the covariance of the Fourier coefficients. Rather than assuming an uninformative flat spectrum, we assign a Lorentzian profile to this covariance. This hyperprior allows the model to infer both the characteristic frequency and the spectral width of the signal. In this way, our approach bridges the gap between purely agnostic searches and template-based models by incorporating the expectation that physical sources typically concentrate their spectral power within a limited frequency range. Additionally, to mitigate potential contamination from unmodeled pulsar noise or transient events, we include pulsar-specific Gaussian processes modeled by Fourier bases at frequencies $k/T$. To maintain computational efficiency, we assign uninformative hyperpriors to each pulsar's Fourier coefficients. These processes help separate incoherent pulsar noise from the coherent individual signal. Our method first identifies the signal's sky location and characteristic frequency spectrum; if the evidence is strong, the signal's shape can then be constructed.

We demonstrate the effectiveness and explore the potential limitations of this approach using four categories of mock datasets, including broad-spectrum and narrow-band signals, a short-duration burst event, and noise-only scenarios. These tests validate our model’s ability to capture diverse signal shapes and reliably distinguish coherent signals from unmodeled pulsar noise.

\section{PTA likelihood including a localized coherent signal\label{likelihood}}

In this section, we first briefly introduce the noise-only PTA likelihood, and then derive the likelihood including a localized coherent signal. Without any knowledge of the potential signal, we model the signal-reduced timing residuals using a Fourier expansion. The Fourier coefficients for the signal can be integrated out analytically. The resulting marginalized likelihood contains only a few parameters that describe the source's sky location and characterize the signal's frequency spectrum.

\subsection{Noise-only likelihood}

Pulsars are rapidly spinning, highly magnetized neutron stars that emit beams of radio waves, sweeping across Earth. Millisecond pulsars are known to have remarkably stable rotations, allowing the times of arrival (TOAs) of the radio pulses to be measured with high precision. Observed TOAs are fit by a timing model that accounts for the pulsar's rotation, astrometry (position/motion), interstellar medium effects, etc. The differences between the measured TOAs and the best-fit TOAs are known as timing residuals. In the absence of deterministic signals, the residuals consist of small perturbations to the timing model, red noise (including pulsar noise and the SGWB), and white noise from measurement uncertainties.

Assuming that the red noise is described by Gaussian processes, the residuals follow a multivariate normal distribution, with log-likelihood \cite{vanHaasteren:2012hj}
\begin{equation}
\log\mathfrak{L}=-\frac{1}{2}\left[\left(r-M\epsilon\right)^{\top}C^{-1}\left(r-M\epsilon\right)+\log\det \left(2\pi C\right)\right],\label{first_L}
\end{equation}
where $r$ is a vector of residuals for all pulsars, $C$ is the combined white and red noise covariance matrix, $M$ is the timing model design matrix, and $\epsilon$ represents small deviations from the best-fit timing model parameters. The covariance matrix $C$ decomposes into
\begin{equation}
C=N+C_{\rm rn},
\end{equation}
where $N$ is the white noise covariance, whose parameters are usually fixed by individual pulsar analyses. The red noise covariance $C_{\rm rn}$ is often approximated in a low-rank form in the frequency domain, truncated at the $n_F$-th Fourier mode:
\begin{equation}
C_{\rm rn}\approx F\phi F^{\top},\label{C_app}
\end{equation}
where $F=\left\{F^{(a)}\right\}$ has a block-diagonal-like structure (one block per pulsar), with each block being an $n^{(a)}_{\rm toa} \times 2n_F$ Fourier design matrix ($n^{(a)}_{\rm toa}$ is the number of TOAs for pulsar $a$):
\begin{equation}
F^{(a)}_{ik} = \left[ \sin\!\left(\frac{2\pi k t^{(a)}_i}{T}\right),\cos\!\left(\frac{2\pi k t^{(a)}_i}{T}\right)\right],\label{Fourier}
\end{equation}
where $t^{(a)}_{i}$ is pulsar $a$'s $i$-th TOA, $k=1,2,\ldots,n_{F}$, and $T$ is the total observation span. In Eq. (\ref{C_app}), $\phi$ contains $n_{\rm psr} \times n_{\rm psr}$ blocks representing pulsar-pair correlations. The block $\phi^{(ab)}$ is given by
\begin{equation}
\phi^{(ab)}_{kl}=\delta_{kl}\left(\delta_{ab}\Phi^{(\rm irn)}_{ak}+\Gamma_{ab}\Phi^{(\rm crn)}_{k}\right),\label{red_noise_block}
\end{equation}
where $a,b$ range over pulsars and $k,l$ over Fourier modes; $\delta_{ij}$ is the Kronecker delta; $\Phi^{(\rm irn)}_{ak}$ describes the spectrum of pulsar $a$'s intrinsic noise; and  $\Gamma_{ab}\Phi^{(\rm crn)}_{k}$ describes processes with a common spectrum across all pulsars and inter-pulsar correlations. For an isotropic SGWB in general relativity, $\Gamma_{ab}$ is given by the Hellings-Downs curve \cite{Hellings:1983fr}. The common red noise component $\Phi^{(\rm crn)}_k$ is typically modeled as a power law,
\begin{equation}
\Phi^{(\rm crn)}_{k}=\frac{A_{\rm crn}^{2}}{12\pi^{2}}\frac{1}{T}\left(\frac{f_{k}}{1\ \text{yr}^{-1}}\right)^{-\gamma_{\rm crn}}\ \text{yr}^{3},\label{eq:power}
\end{equation}
where $A_{\rm crn}$ is the amplitude and $\gamma_{\rm crn}$ is the spectral index. For an SWGB generated by inspiraling SMBHBs, $\gamma_{\rm crn}$ is expected to be $13/3$ \cite{phinney2001practical}. The intrinsic pulsar noise $\Phi_{ak}^{(\rm irn)}$ is usually also modeled as a power law, with pulsar-specific amplitudes and spectral indices.

Besides the common and intrinsic red noise, individual pulsars may exhibit additional unmodeled noise or transient events. The low-rank approximation in Eq.~(\ref{C_app}) can equivalently be viewed as modeling the red noise through a Fourier expansion, with priors on the Fourier coefficients set by a multivariate normal distribution with covariance $\phi$. This same formalism can therefore be used to account for unmodeled noise (un). For simplicity, we assume that all Fourier coefficients have the same variance:
\begin{equation}
    \Phi^{(\rm un)}_{ak} = \frac{A_{\mathrm{un}}^{(a)2}}{12\pi^{2}}\frac{1}{T}\ \text{yr}^{3},\label{eq:un}
\end{equation}
where $A_{\mathrm{un}}^{(a)}$ characterizes the noise amplitude of pulsar $a$. This corresponds to a flat spectrum, as it is a special case of a power law spectrum with a vanishing spectral index. A similar spectrum was introduced in Ref. \cite{NANOGrav:2023gor} to handle the unmodeled white noise at high frequencies.

In standard practice, the likelihood (\ref{first_L}) is further simplified by marginalizing over the linear perturbations to the timing model. Since the TOAs are dominated by the timing model, we assign uninformative priors to the timing model parameters and, therefore, also to the first order perturbations $\epsilon$. This allows us to integrate out $\epsilon$ analytically, which effectively projects the residuals into the subspace orthogonal to the timing model design matrix $M$. In other words, residuals induced by any GW signals, whether stochastic or deterministic, are automatically pre-processed with their ``timing model components'' removed. Only the post-fit residuals are relevant for the likelihood.

Using the Woodbury identity,  the marginalized PTA likelihood $\mathcal{L}=\int\mathfrak{L}\mathrm{d}\epsilon$ is
\begin{equation}
\log{\mathcal{L}}  =-\frac{1}{2}\left(r^{\top}K^{-1}r+\log\det K\right)+\ldots, \label{eq:Lapprox}
\end{equation}
where ``$\ldots$'' denotes constant terms irrelevant to Bayesian inference. Here,
\begin{equation}
K^{-1}=D^{-1}-D^{-1}F\left(\phi^{-1}+ F^{\top}D^{-1}F\right)^{-1} F^{\top}D^{-1},
\end{equation}
with
\begin{equation}
D^{-1}=N^{-1}-N^{-1}M\left(M^{\top}N^{-1}M\right)^{-1}M^{\top}N^{-1}. \label{eq:D}
\end{equation}
Eq. (\ref{eq:Lapprox}) is the likelihood implemented in production-level GW search pipelines, such as \texttt{Enterprise} \cite{enterprise,enterprise_extension} and \texttt{Discovery} \cite{Vallisneri_nanograv_discovery_2025}.

\subsection{Timing residuals induced by a localized deterministic signal}

A plane gravitational wave (GW) contributes to the timing residuals through the pulsar term and the Earth term \cite{estabrook1975response}. The pulsar terms are generally poorly correlated between different pulsars, since their phases depend on pulsar distances, which are often not well constrained. Moreover, for non-continuous GW signals, the pulsar-term contributions from different pulsars are separated in time by hundreds to thousands of years. Even if the pulsar distances are precisely known, since the pulsar terms are sensitive to the long-term evolution of the signal, it is hard to model the waveform in an agnostic way.  We therefore focus on the Earth term, which induces a simultaneous signal in the timing residuals of all pulsars, with correlations fully determined by the sky locations of the source and pulsars, as well as the GW polarization.

For a localized deterministic GW signal with strains $h_{+/\times}(t)$, the induced residual at time $t$ is:
\begin{equation}
s(t)= F_{+}(\hat{\Omega})\int_{t_0}^{t} h_{+}(t^\prime)\text{d}t^\prime+F_{\times}(\hat{\Omega})\int_{t_0}^{t} h_{\times}(t^\prime)\text{d}t^\prime.\label{s(t)}
\end{equation}
where $\hat{\Omega}$ denotes the unit vector pointing from the GW source toward the Solar System barycenter, and $t_0$ is the beginning of the observation span. $F_{+/\times}(\hat{\Omega})$ are the antenna pattern functions that describe the geometric projection between the pulsar line of sight and the GW propagation direction. For a specific waveform template, $s$ is determined by the waveform parameters or the underlying physical parameters. A few examples are presented in Appendix \ref{append}. In this work, however, we are interested in the more general case where the form of $h_{+/\times}(t)$ is unknown.

\subsection{Likelihood with a localized signal}

When a localized deterministic signal is present, the residuals $r$ in the noise-only likelihood (\ref{eq:Lapprox}) should be replaced by ${r}-s$, where $s=\left\{s^{(a)}\right\}$, given by Eq. (\ref{s(t)}), denotes the signal-induced residuals. Here, being agnostic about what is possibly buried in the noise, we model $s^{(a)}$ of pulsar $a$ using a Fourier expansion:
\begin{equation}
s^{(a)} \approx F_{+}^{(a)}P^{(a)}H_{+} +F_{\times}^{(a)}P^{(a)}H_{\times}, \label{FP}
\end{equation}
where $H_{+/\times}$ are vectors of Fourier coefficients shared by all pulsars, and $P^{(a)}$ is a Fourier design matrix,
\begin{equation}
P^{(a)}_{ik} = \left[ \sin\left(\frac{2\pi k t^{(a)}_i}{T_P}\right),\cos\!\left(\frac{2\pi k t^{(a)}_i}{T_P}\right)\right],
\end{equation}
with $k=1,2,\ldots,n_{P}$. This has the same form as Eq. (\ref{Fourier}), but may have a different cutoff $n_P$ and a different frequency grid determined by $T_P$. Because Fourier modes at frequencies $1/T,2/T,\ldots$ result in a periodic function of period $T$, choosing $T_P=T$ can lead to boundary oscillations (the Gibbs phenomenon). Extending the span mitigates this effect by allowing the reconstructed signal to be non-periodic.\footnote{This was also proposed in Ref. \cite{vanHaasteren:2014faa} to handle stationary noise processes such as the SGWB.} In this work, we take $T_P = 3T$, i.e., $T_P$ is three times the PTA observation span. 

For notational convenience, we define an $n^{(a)}_{\rm toa}\times4n_{{P}}$ matrix $S^{(a)}=\begin{pmatrix}F_{+}^{(a)}P^{(a)} & F_{\times}^{(a)}P^{(a)}\end{pmatrix}$. The signal-induced residuals $s=\left\{s^{(a)}\right\}$ can then be written as
\begin{equation}
    s=SH,
\end{equation}
where $S=\begin{pmatrix}{S}^{(1)\top} &{S}^{(2)\top} & \cdots & {S}^{(n_{\text{psr}})\top}\end{pmatrix}^{\top}$ and $H=\begin{pmatrix}H_+ & H_\times\end{pmatrix}^{\top}$. 

Define the inner product $\left< x|y\right> =x^{\top}K^{-1}y$. The likelihood becomes
\begin{align}
\log \mathcal{L}&=-\frac{1}{2}\left(\langle r-SH|r-SH\rangle+\log\det K\right)\\
&=-\frac{1}{2}\left(\langle r|r\rangle+\log\det K\right)
 + \langle r|S\rangle H-\frac{1}{2}H^\top\langle S|S\rangle H,\label{L}
\end{align}
where the first term in the second line is simply the noise-only likelihood (Eq. (\ref{eq:Lapprox})). The sky location of the source $(\theta,\varphi)$ appears in the two inner products $\left<r|S\right>$ and $\left<S|S\right>$. 

To perform Bayesian inference, we must specify priors for the Fourier coefficients $H$.  When searching for physically motivated signals (e.g., SMBHB continuous waves), Fourier coefficients typically do not have simple priors because they depend nonlinearly on parameters (e.g., the chirp mass). Here, since we are agnostic about the waveform, we use a Gaussian prior $\pi(H|Q)$:
\begin{equation}
    \log \pi(H|Q)=-\frac{1}{2}\left[H^\top Q^{-1} H + \log\det \left(2\pi Q\right)\label{pi_H}\right]
\end{equation}
where, for simplicity, the covariance $Q$ is assumed to be diagonal. Each diagonal entry corresponds to a sine or cosine component at frequency $f_k$, so the diagonal can be interpreted as the ``spectrum'' of the induced residuals. As discussed in the previous subsection, for unmodeled pulsar noise transients, we adopt a flat spectrum (Eq. (\ref{eq:un})), which assigns one prior to all Fourier coefficients. This is the simplest choice, involving only a single hyperparameter. For the localized signal, however, we want the spectrum to be more informative. We choose
\begin{equation}
    Q=\mathrm{diag}\{Q_+, Q_\times\},
\end{equation}
with 
\begin{equation}
    Q_{+/\times}(f_k)=\frac{q^2_{+/\times}}{1+\left(\frac{f_k-f_c}{f_w}\right)^2},\label{cauchy}
\end{equation}
which has the shape of a Lorentzian function with central frequency $f_c$ and width $f_w$. This choice reflects the expectation that a GW signal is typically concentrated around a characteristic frequency set by the source’s dynamical timescale, which in turn is determined by its mass and size. The parameters $f_c$ and $f_w$ control the dominant frequency and the frequency range, respectively. For example, for continuous waves, $f_c$ corresponds approximately to the orbital frequency of the SMBHB, and $f_w$ is small since the signal is dominated by a narrow frequency band. This Lorentzian profile is also able to cover a broad frequency range: as $f_w\to\infty$, $Q\to q^2$, and the spectrum reduces to a flat one. Compared with, e.g., a Gaussian profile, the Lorentzian shape has heavy tails, providing conservative support at the frequency ends. Finally, we also assume that the two polarization modes $H_{+/\times}$ share the same spectral shape but can have different amplitudes. Thus, the spectrum is governed by only four hyperparameters: $q_+, q_\times, f_c$ and $f_w$.

Given the Gaussian prior (\ref{pi_H}), we can analytically integrate out the Fourier coefficients $H$ in Eq. (\ref{L}) and obtain the marginalized likelihood $L=\int\mathcal{L}\mathrm{d}H$:
\begin{empheq}[box=\boxed]{equation}
\begin{aligned}
\log L &=-\frac{1}{2}\left(\langle r|r \rangle + \log\det K\right)\\
&\quad + \frac{1}{2}\left(\langle r|S \rangle \Sigma^{-1} \langle S|r \rangle
       - \log\det Q\Sigma\right)
\end{aligned}
\label{model_likelihood}
\end{empheq}
where $\Sigma=\left<S|S\right>+Q^{-1}$. This is the likelihood for our Bayesian inference. It has the same form as the likelihood derived in Ref. \cite{Deng:2023btv}, where the matrix $S$ encodes the linear interpolation design matrix, while here it contains the Fourier basis. Parameters entering the Bayesian inference include the SGWB parameters such as $A_{\rm crn}$ and $\gamma_{\rm crn}$, pulsar noise parameters such as $A_{\text{un}}^{(a)}$, the sky location angles $\theta$ and $\varphi$, and the hyperparameters $q_+, q_\times, f_c$ and $f_w$ that characterize the signal spectrum.

\subsubsection*{Signal reconstruction\label{subsec:Waveform-reconstruction}}

If our waveform-agnostic model is favored over the noise-only model, one would be interested in what the signal looks like. The signal-induced residuals can be reconstructed straightforwardly from posterior samples. From Eqs. (\ref{L}) and (\ref{pi_H}), the conditional posterior of the Fourier coefficients $H$ is a normal distribution with covariance $\Sigma^{-1}$ and mean $\Sigma^{-1}\left<S|r\right>$: $p(H)\sim \mathcal{N}(\Sigma^{-1}\left<S|r\right>,\Sigma^{-1})$. To obtain the marginal posterior $p(H)$, we need to marginalize over the other model parameters. Using samples obtained based on the likelihood (\ref{model_likelihood}), we have
\begin{equation}
    p(H)\propto \sum_j \mathcal{N}\left(\Sigma_{(j)}^{-1}\left<S_{(j)}|r\right>,\Sigma_{(j)}^{-1}\right),\label{eq:reconstruct}
\end{equation}
where $j$ labels the $j$-th posterior sample. In practice, for each posterior sample, we draw one realization of $H$ from $\mathcal{N}\left(\Sigma_{(j)}^{-1}\left<S_{(j)}|r\right>,\Sigma_{(j)}^{-1}\right)$. The resulting samples of $H$ can then be used to reconstruct the signal shape via a Fourier expansion.
\\
\par
In summary, we model the residuals induced by an agnostic individual signal using a Fourier expansion, whose coefficients are analytically integrated out. The resulting marginalized likelihood takes a simple form, and the signal parameters are reduced to the sky location angles $(\theta,\varphi)$ and four hyperparameters $q_+, q_\times, f_c$ and $f_w$ that characterize the frequency-domain profile of the induced residuals. We also add parameters $A_{\text{un}}^{(a)}$ to each pulsar to capture unmodeled noise or transients. If a signal is detected, the induced residuals can be reconstructed with the help of Eq.~(\ref{eq:reconstruct}).

\begin{table*}[ht]
\begin{center}
\caption{Parameters and their injected values in mock datasets. The prior column lists the priors of the SGWB parameters $A_{\rm crn}$ and $\gamma_{\rm crn}$ (used for analyses in all datasets), and the priors of the parameters for the burst signal (used in the waveform-template model for dataset BURST).}
\label{tab:injected}
\begin{tabular}{llll}
\hline\hline
{Parameter}~~~~~ & {Description} & {Injected value}  & {Prior}\\
\hline

\multicolumn{4}{c}{\textbf{\textit{Power-law CURN (all datasets)}}} \\[1pt]
$A_{\mathrm{crn}}$ & Amplitude & $5\times10^{-14}$ & log-uniform $[-18, -11]$\\
$\gamma_{\rm crn}$& Spectral index & $13/3$ & uniform $[0, 7]$\\
\hline

\multicolumn{4}{c}{\textbf{\textit{Sinc signal (SINC)}}} \\[1pt]
$\cos\theta_{\rm sc}$ & Polar angle  & $0.5$ & \\ 
$\varphi_{\rm sc}$ & Azimuthal angle  & $4.5$& \\ 
$A_{\rm sc}$ [s] & Amplitude  & $10^{-5.6}$& \\ 
$f_{\rm sc}$ [Hz] & Frequency  & $5\times 10^{-8}$ \\ 
$t_{\rm sc}$ [s] & Time when sinc peaks & $4.7\times 10^9$ & \\
$\psi_{\rm sc}$ & Polarization angle & $0$ & \\
\hline

\multicolumn{4}{c}{\textbf{\textit{Sinusoidal signal (SINE)}}} \\[1pt]
$\cos\theta_{\rm s}$ & Polar angle  & $0.5$\\ 
$\varphi_{\rm s}$ & Azimuthal angle  & $4.5$\\ 
$A_{\rm s}$ [s] & Amplitude  & $10^{-7}$\\ 
$f_{\rm s}$ [Hz] & Frequency  & $2\times 10^{-8}$ \\ 
$\phi_{\rm s}$ & Phase & $1$  \\
$\psi_{\rm s}$ & Polarization angle & $0$\\
\hline

\multicolumn{4}{c}{\textbf{\textit{Short burst (BURST)}}} \\[1pt]
$\cos\theta_{\rm b}$ & Polar angle  & $0$ & uniform $[-1, 1]$\\ 
$\varphi_{\rm b}$ & Azimuthal angle  & $\pi$& uniform $[0, 2\pi]$\\ 
$A_{\rm b}$ [s] & Amplitude  & $10^{-5.78}$& log-uniform $[-10, -5]$\\ 
$\sigma_{\rm b}$ [s] & Duration  & $5\times 10^6$& log-uniform $[6.2, 8.5]$ \\
$t_{\rm b}$ [s] & Time when burst peaks & $4.7\times 10^9$ & uniform $[4.58\times10^9, 4.89\times10^9]$\\
$\psi_{\rm b}$ & Polarization angle & $0$ & uniform $[0, 2\pi]$\\
\hline

\multicolumn{4}{c}{\textbf{\textit{Sine-Gaussian pulsar noise (SINE, D1 and D2)}}} \\[1pt]
$A_{\mathrm{sg}}^{(a)}$ [s]& Gaussian amplitude & $10^{-6.2}$\\
$t_{\mathrm{sg}}^{(a)}$ [s]& Gaussian mean & $4.75\times10^{9}$\\
$\sigma_{\mathrm{sg}}^{(a)}$ [s]& Gaussian width & $3\times10^7$\\
$f_{\mathrm{sg}}^{(a)}$ [$\text{Hz}$]& Sinusoidal frequency & $10^{-7.4}$ (SINE, D1)\\
&& Drawn from log-uniform [-7.6,-7] (D2)\\
$\phi_{\mathrm{sg}}^{(a)}$ & Sinusoidal phase & $3$\\
\hline

\end{tabular}
\end{center}
\end{table*}

\begin{table*}[ht]
\begin{center}
\caption{Priors for waveform-agnostic model}
\label{tab:priors}
\begin{tabular}{llll}
\hline\hline
{Parameter}~~~~~ & {Description}~~~~~~~~~~~~~~~ & {Prior}\\
\hline

\multicolumn{3}{c}{\textbf{\textit{Unmodeled pulsar noise (flat spectrum)}}} \\[1pt]
$A_{\mathrm{un}}^{(a)}$ & Amplitude & log-uniform $[-18, -11]$\\
\hline

\multicolumn{3}{c}{\textbf{\textit{Source sky location}}} \\[1pt]
$\cos\theta$& Polar angle & uniform $[-1, 1]$\\
$\varphi$& Azimuthal angle & uniform $[0, 2\pi]$\\
\hline

\multicolumn{3}{c}{\textbf{\textit{Signal spectrum (Lorentzian shape, $T=10~\text{yr}$)}}} \\[1pt]
$q_{+/\times}$ [s] & Amplitudes & log-uniform $[-10, -5]$\\
$f_c\times 3T$& Central frequency & uniform $[1,
 60]$\\
$f_w\times 3T$& Width & log-uniform $[\log_{10}0.3,\log_{10}120]$\\

\hline

\end{tabular}
\end{center}
\end{table*}

\section{Analyses of mock datasets \label{simulated data}}

We test our model by analyzing five mock datasets. Each dataset consists of twenty pulsars with an observation span of ten years and a cadence of fifteen days. For simplicity, all residuals have the same constant white noise level of $0.5\ \mu$s, and no intrinsic red noise. In addition, an SGWB is injected with a power law spectrum given by Eq. (\ref{eq:power}), with $A_{\rm crn}=4\times10^{-15}$ ($\log_{10}A_{\rm crn}\approx-14.398$) and $\gamma_{\rm crn}=13/3$. For computational efficiency, we treat the background as a common uncorrelated red noise (CURN) process, which means $\Gamma_{ab}=\delta_{ab}$ in Eq. (\ref{red_noise_block}). This model, commonly adopted by PTAs when the inter-pulsar correlations are not significant, greatly reduces the computational cost because the noise matrix $C$ is now block-diagonal, allowing its inverse to be computed block by block (or pulsar by pulsar).

We generate five datasets to test the capabilities and potential limitations of the model:
\begin{enumerate}
    \item Dataset SINC (\textit{broad-spectrum signal}). This dataset contains a strong signal whose induced residuals are described by a sinc function, $\sin(x)/x$. Such a signal does not come from any known physical processes; we use it because the sinc function's Fourier transform is flat and broad. No pulsar-specific noise transients are added. We aim to demonstrate that our waveform-agnostic model can correctly recover the signal, including its broad spectrum.
    
\item Dataset SINE (\textit{narrow-spectrum signal}). This dataset contains a moderate signal whose induced residuals are described by a sinusoid. Such a signal could arise from an inspiraling SMBHB. The Fourier transform of a sinusoid is sharply localized at the sinusoidal frequency. In addition to the signal, we add sine-Gaussian noise transients to five pulsars. We aim to correctly recover the signal and test whether the pulsar noise can be distinguished from the coherent signal.

    \item Dataset BURST (\textit{comparison with waveform-template model}). This dataset contains a short burst signal. No pulsar-specific noise transients are added. We aim to compare the waveform-agnostic model with the true model. In the true model, the search uses the same waveform that generated the residuals; it is therefore, by construction, the optimal model for the data.

\item Datasets D1 and D2 (\textit{possible false alarm}). The fourth dataset, D1, contains no coherent signals, but has identical sine-Gaussian noise transients in ten pulsars. The purpose is to test whether the incoherent pulsar noise can be misidentified as a coherent signal in this contrived scenario. The fifth dataset, D2, differs from D1 only in the sine-Gaussian transients present in the ten pulsars. Rather than being identical, the ten transients, occurring at the same epoch, have different frequencies. The purpose, again, is to test whether the pulsar noise can be misidentified as a coherent signal by our model.
\end{enumerate}

In all tests, we set the high-frequency cutoff of the common red noise to $20/T$, where $T$ is ten years. In the waveform-agnostic model, we set the same cutoff for the signal, so the Fourier basis has $60$ sine-cosine components because the discrete frequencies are at $1/T_P, 2/T_P,\ldots$ with $T_P=3T$. Unless otherwise specified, all Bayesian searches are carried out by \texttt{Enterprise} and \texttt{PTMCMCSampler} \cite{justin_ellis_2017_1037579}.

\subsection{Strong sinc signal without pulsar noise transients}
In the first dataset, we inject signal-induced residuals described by a sinc function ($\text{sinc}(x)\equiv \sin(x)/x$). For pulsar $a$, the induced residuals are
\begin{align}
    r^{(a)}(t) =&~ \left[F^{(a)}_+(\theta_{\rm sc}, \varphi_{\rm sc})\cos\psi_{\rm sc}+F^{(a)}_{\times}(\theta_{\rm sc}, \varphi_{\rm sc})\sin\psi_{\rm sc}\right]\notag\\&\times A_{\rm sc} \frac{\sin\left[2\pi f_{\rm sc} \left(t-t_{\rm sc}\right)\right]}{2\pi f_{\rm sc} \left(t-t_{\rm sc}\right)}.
\end{align}
Here, $\psi_{\rm sc}$ is the polarization angle (we assume linear polarization), $A_{\rm sc}$ is the signal amplitude, and $t_{\rm sc}$ denotes the time when the sinc function reaches its peak. The injected parameter values are listed in Table \ref{tab:injected}. Although such a signal is not associated with any known physical process, we adopt it because the sinc function has an interesting spectrum: the Fourier transform of $\sin(2\pi f_{\rm sc} t)/t$ has constant support within $[-f_{\rm sc},f_{\rm sc}]$ and vanishes outside this interval. This rectangular frequency profile can be roughly described by a Lorentzian function with a central frequency $f_c\sim 0$ and width $f_w \sim f_{\rm sc}$. Additionally, we inject a signal with a relatively large amplitude $A_{\rm sc}$, several times larger than the white noise level, in order to test the waveform-agnostic model's ability to reconstruct the injected residuals.

\begin{figure}[h!]
\centering
\includegraphics[scale=0.35]{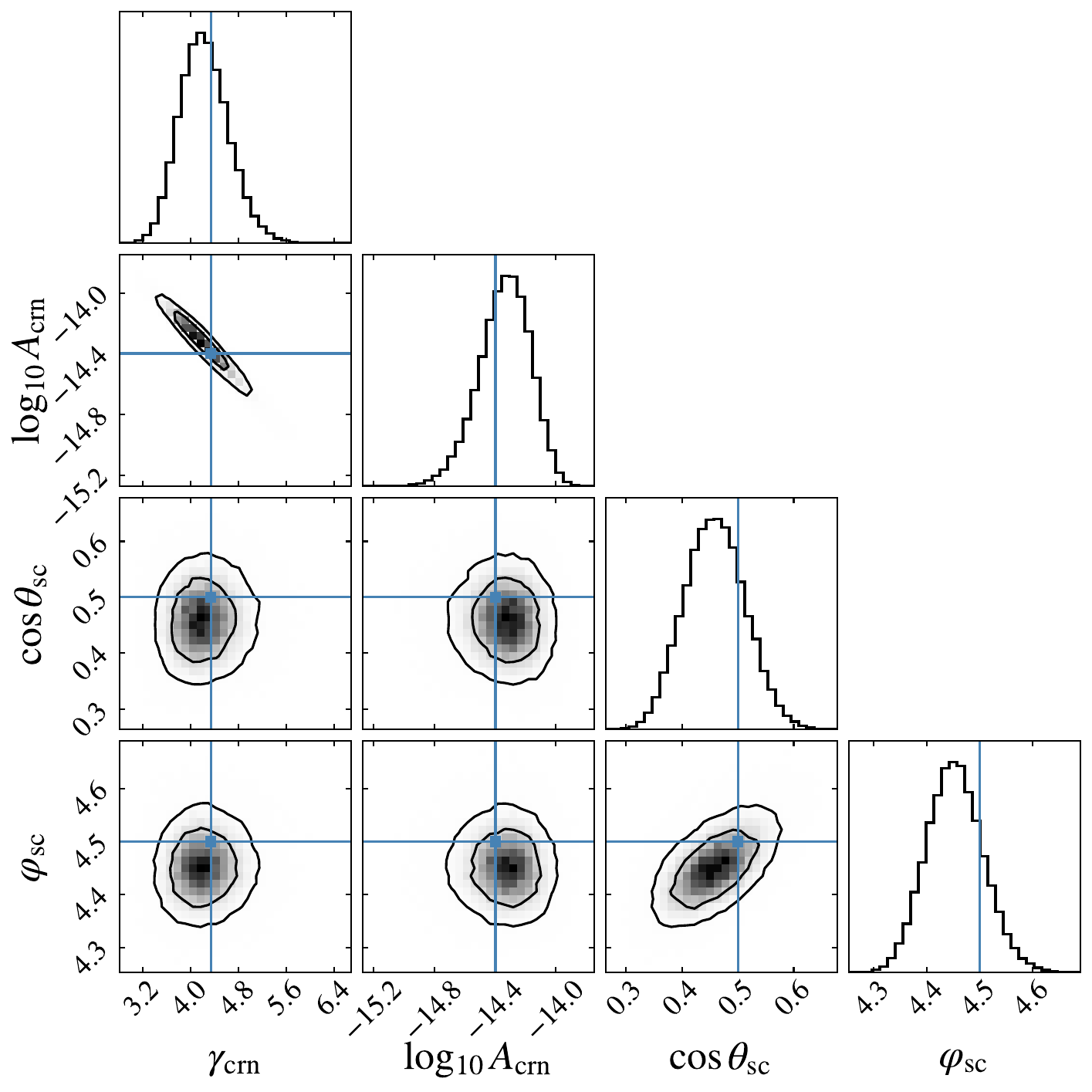}

\caption{\label{fig:sinc_corner}Corner plot of $\gamma_{\rm crn}, \log_{10}A_{\rm crn}, \cos\theta_{\rm sc}$ and $\varphi_{\rm sc}$ in the waveform-agnostic model for dataset SINC. The blue lines show the injected parameter values. The model accurately localizes the source and recovers the SGWB.}
\end{figure}

\begin{figure}[h!]
\centering
\includegraphics[scale=0.35]{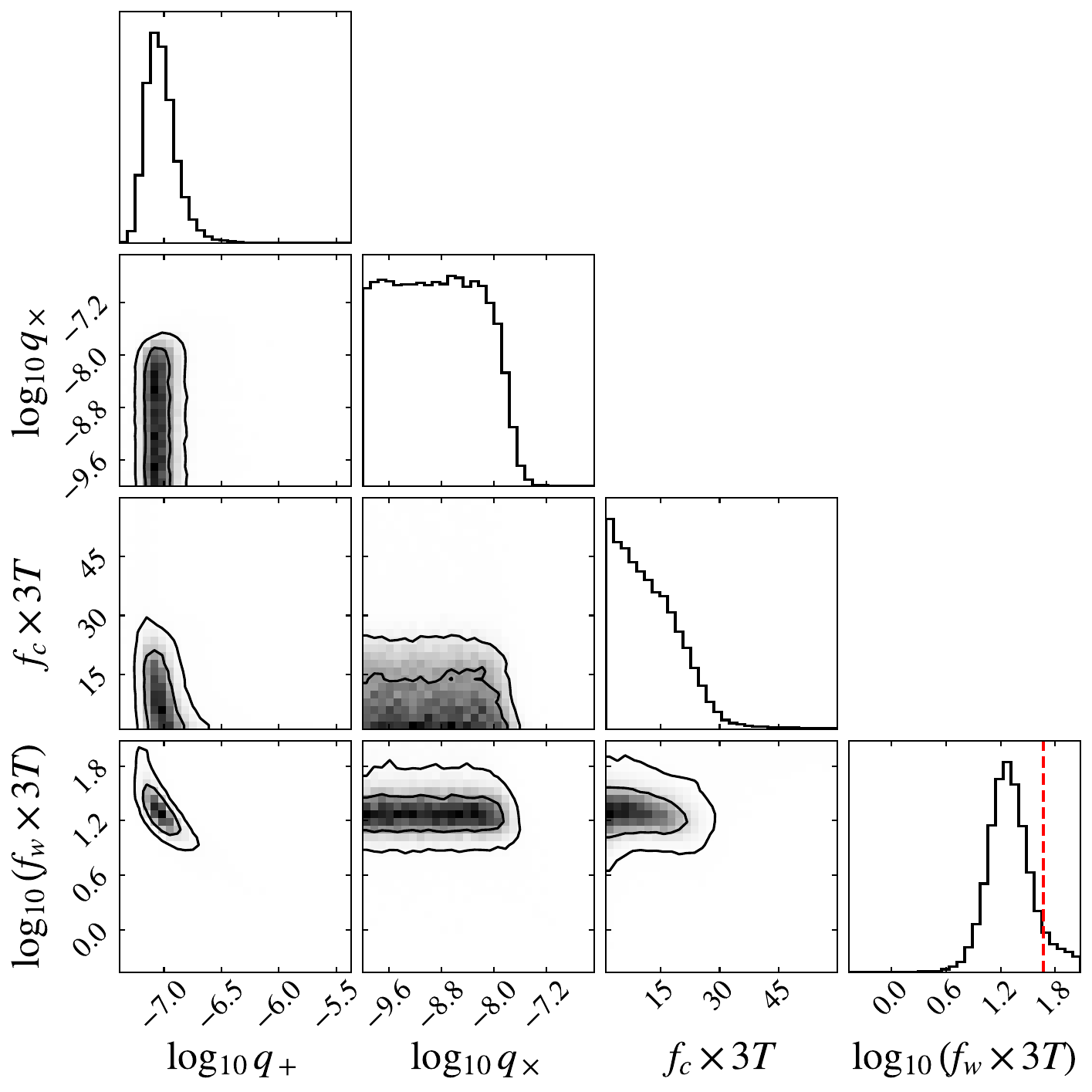}

\caption{\label{fig:sinc_q+qcross} Corner plot of $\log_{10}q_+, \log_{10}q_\times,f_c$ and $\log_{10}f_w$ for dataset SINC. A strong plus polarization mode is detected. The red dashed line represents the frequency of the injected sinc signal $f_{sc}$. The Fourier transform of the sinc function has constant support within $[-f_{\rm sc}, f_{\rm sc}]$, which is indeed captured by $f_w$ and $f_c$.}
\end{figure}

Analysis of the dataset gives the following results. Fig. \ref{fig:sinc_corner} shows the corner plot of $\gamma_{\rm crn}, \log_{10}A_{\rm crn}, \cos\theta_{\rm sc}$ and $\varphi_{\rm sc}$. All injected features are accurately captured, including a tight constraint on the source's sky location. The parameters of the Lorentzian spectrum are shown in Fig. \ref{fig:sinc_q+qcross}. The posterior of $\log_{10}q_+$ has strong support, whereas the posterior of $\log_{10}q_\times$ indicates non-detection, consistent with the injected signal, which has $\psi_{\rm sc}=0$ and a significant signal amplitude $A_{sc}$. Furthermore, the posterior of $f_c$ favors the lower prior bound, and the posterior of $f_w$ peaks at $10^{1.3}/(3\times 10~ \text{yr})\approx 2\times 10^{-8}~ \text{Hz} \sim f_{\rm sc}=5\times 10^{-8} ~\text{Hz}$. These suggest that the underlying rectangular frequency profile is well captured by the Lorentzian profile.

\begin{figure}[h!]
\centering
\includegraphics[scale=0.35]{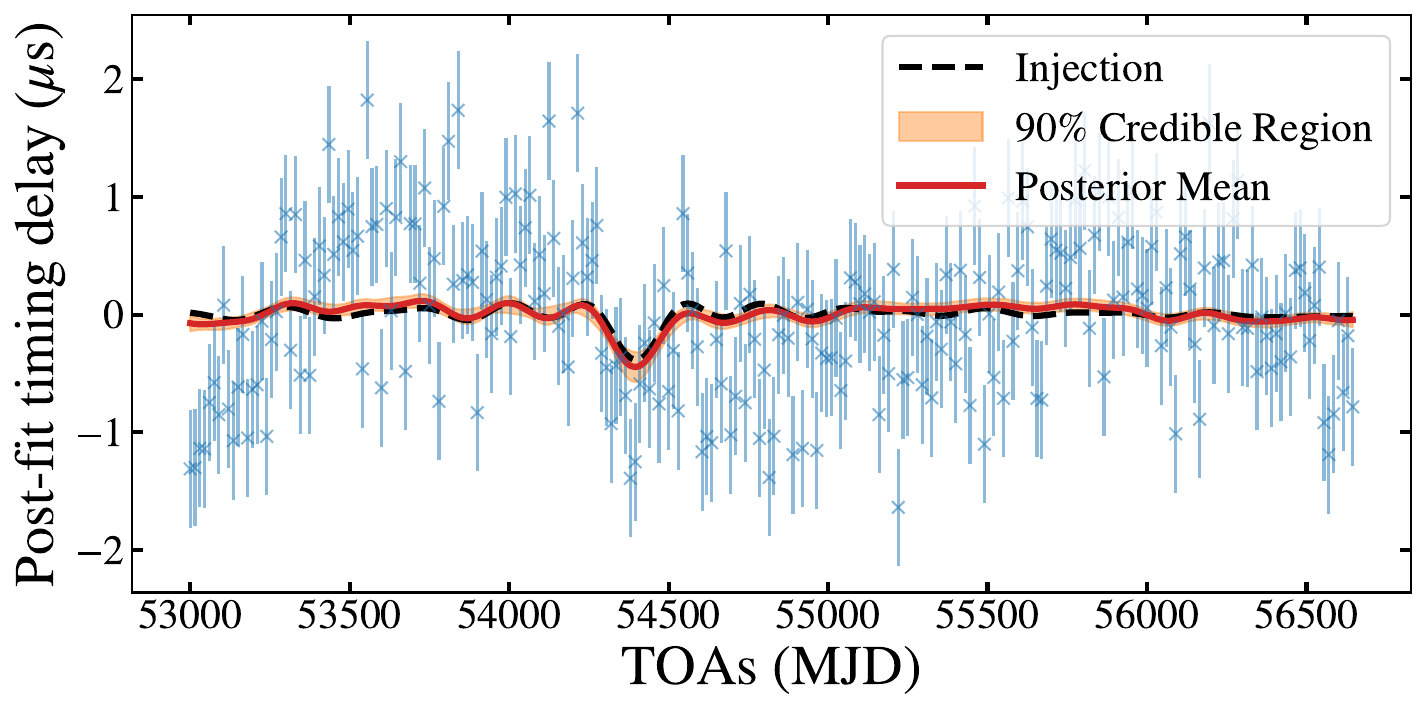}
\includegraphics[scale=0.35]{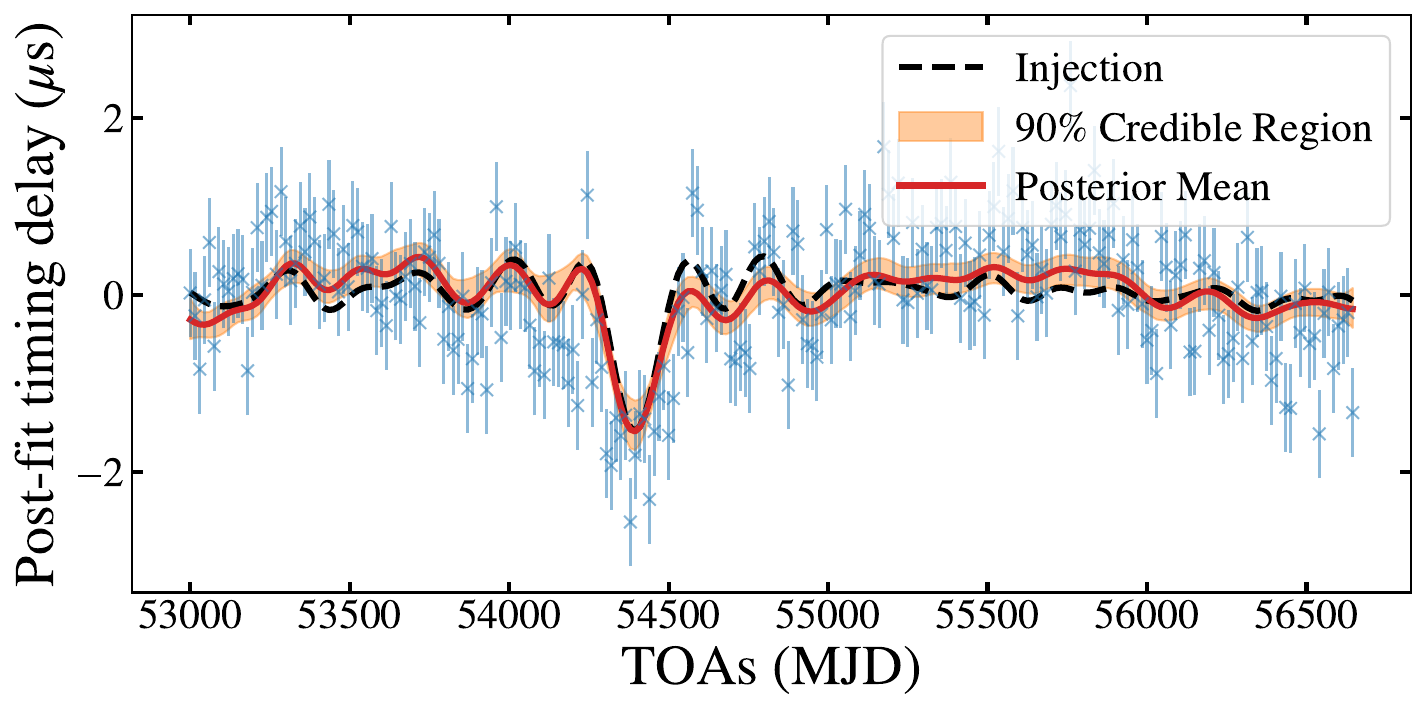}
\includegraphics[scale=0.35]{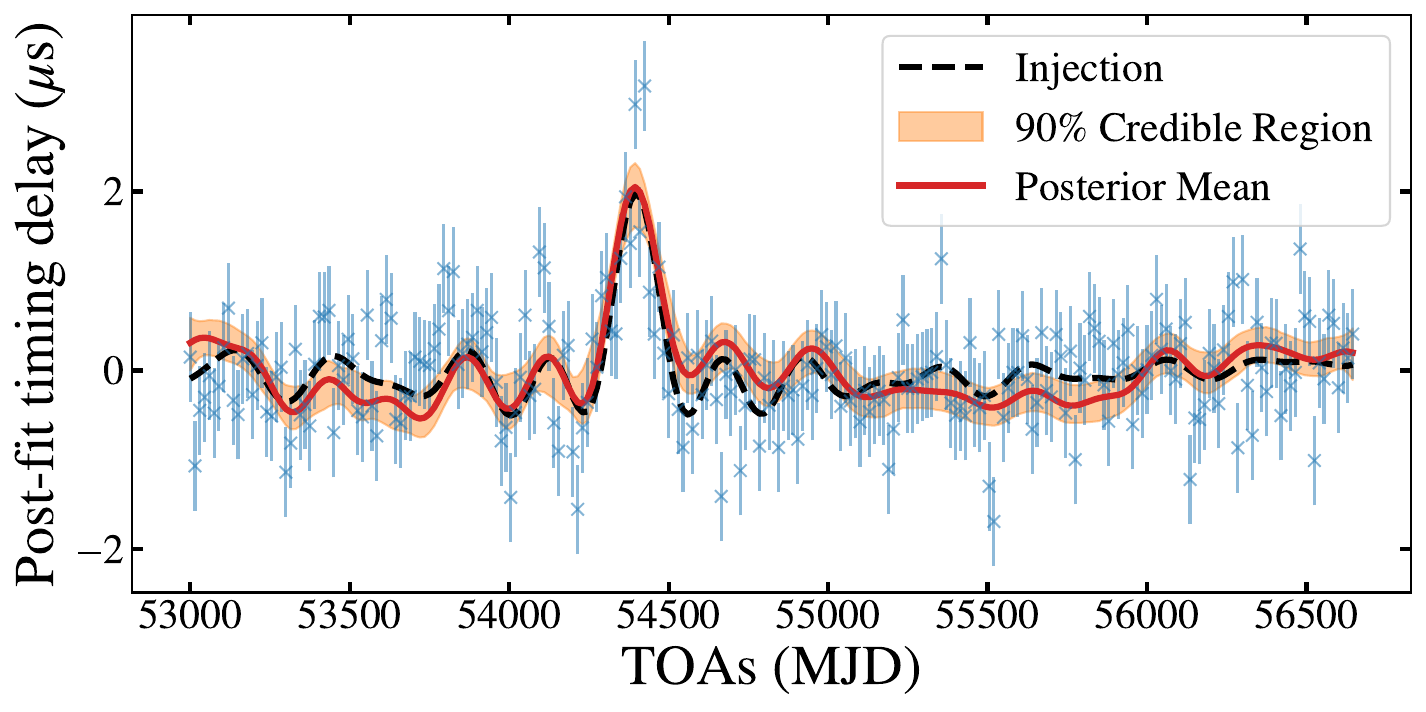}
\caption{\label{fig:sinc_reconstruction}Reconstruction of the signal in three pulsars (Pulsars 0, 2 and 8 from the top) in the waveform-agnostic model for dataset SINC. The blue error bars represent the timing residuals with measurement uncertainties.}
\end{figure}

The reconstruction of the signal-induced residuals in three pulsars (Pulsars 0, 2, and 8) is shown in Fig. \ref{fig:sinc_reconstruction}. Here, all timing delays, including the residuals, injected signals, and reconstructed signals, are "post-fit". We can see that the injections (black) are faithfully reconstructed (red) with narrow credible bands. This experiment demonstrates that the waveform-agnostic model is able to detect and correctly characterize a sufficiently strong deterministic signal.

\begin{figure}[h!]
\centering
\includegraphics[scale=0.4]{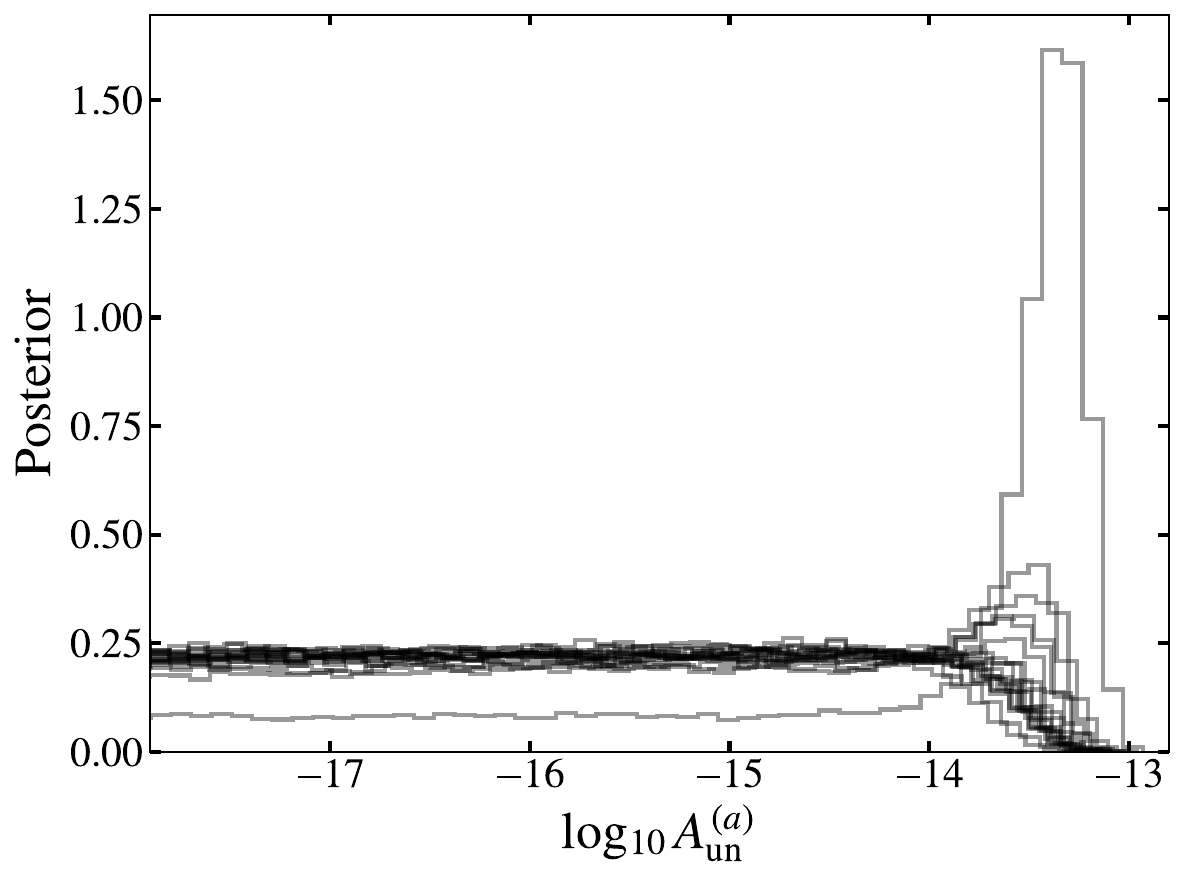}

\caption{\label{fig:sinc_A} Posteriors of $\log_{10}A^{(a)}_{\rm un}$ for the twenty pulsars in the waveform-agnostic model for dataset SINC.}
\end{figure}

Lastly, in Fig. \ref{fig:sinc_A} we show the posteriors of $\log_{10}A^{(a)}_{\rm un}$, which represent the amplitudes of the flat spectra introduced to absorb potential unmodeled pulsar-specific noise transients. As expected, since no individual transients are added, most of the posteriors are simple plateaus, indicating non-detections. Nevertheless, we observe a distinct peak for one pulsar (Pulsar 2), which implies that the flat spectra may have undesirably stolen some power from the deterministic signal (or the SGWB), thus biasing the inference. By the Savage-Dickey ratio, the Bayes factor favoring the existence of a transient in Pulsar 2 is only $\sim 1.9$.

\subsection{Moderate sinusoidal signal with pulsar noise transients in five pulsars \label{sine}}

The second dataset contains a localized, linearly polarized sinusoidal signal. The signal is intended to mimic continuous waves from an Earth-term-only, edge-on SMBHB. Compared with the sinc signal that has a wide and flat spectrum, a sinusoid has only one frequency. The induced residuals in pulsar $a$ are given by
\begin{align}
    r^{(a)}(t) =&~ \left[F^{(a)}_+(\theta_{\rm s}, \varphi_{\rm s})\cos\psi_{\rm s}+F^{(a)}_{\times}(\theta_{\rm s}, \varphi_{\rm s})\sin\psi_{\rm s}\right]\notag\\&\times A_{\rm s} \sin(2\pi f_{\rm s}t+\phi_{\rm s}).
\end{align}
The injected parameter values are listed in Table \ref{tab:injected}.  The signal amplitude is chosen to be $0.16~\mu\text{s}$, well below the white noise level $0.5~\mu\text{s}$. In order to test whether the waveform-agnostic model can distinguish the coherent signal from the incoherent pulsar noise, we add sine-Gaussian transients
\begin{equation}
    r_{\rm sg}(t) = A_{\rm sg}\sin\left[2\pi f_{\rm sg} (t - t_{\rm sg}) + \phi_{\rm sg}\right] e^{-\frac{(t - t_{\rm sg})^2}{2\sigma_{\rm sg}^2} }
\end{equation}
to five pulsars. To further ``confuse'' the model, the five transients are identical, with parameter values listed in Table \ref{tab:injected}. Their amplitude is comparable to the white noise level. In reality, five pulsars having exactly the same noise transients is unlikely. The setup is designed to test the model's ability to identify the coherent signal by adjusting the antenna pattern. The priors of the model parameters are listed in Tables \ref{tab:injected} and \ref{tab:priors}.

\begin{figure}[h!]
\centering
\includegraphics[scale=0.35]{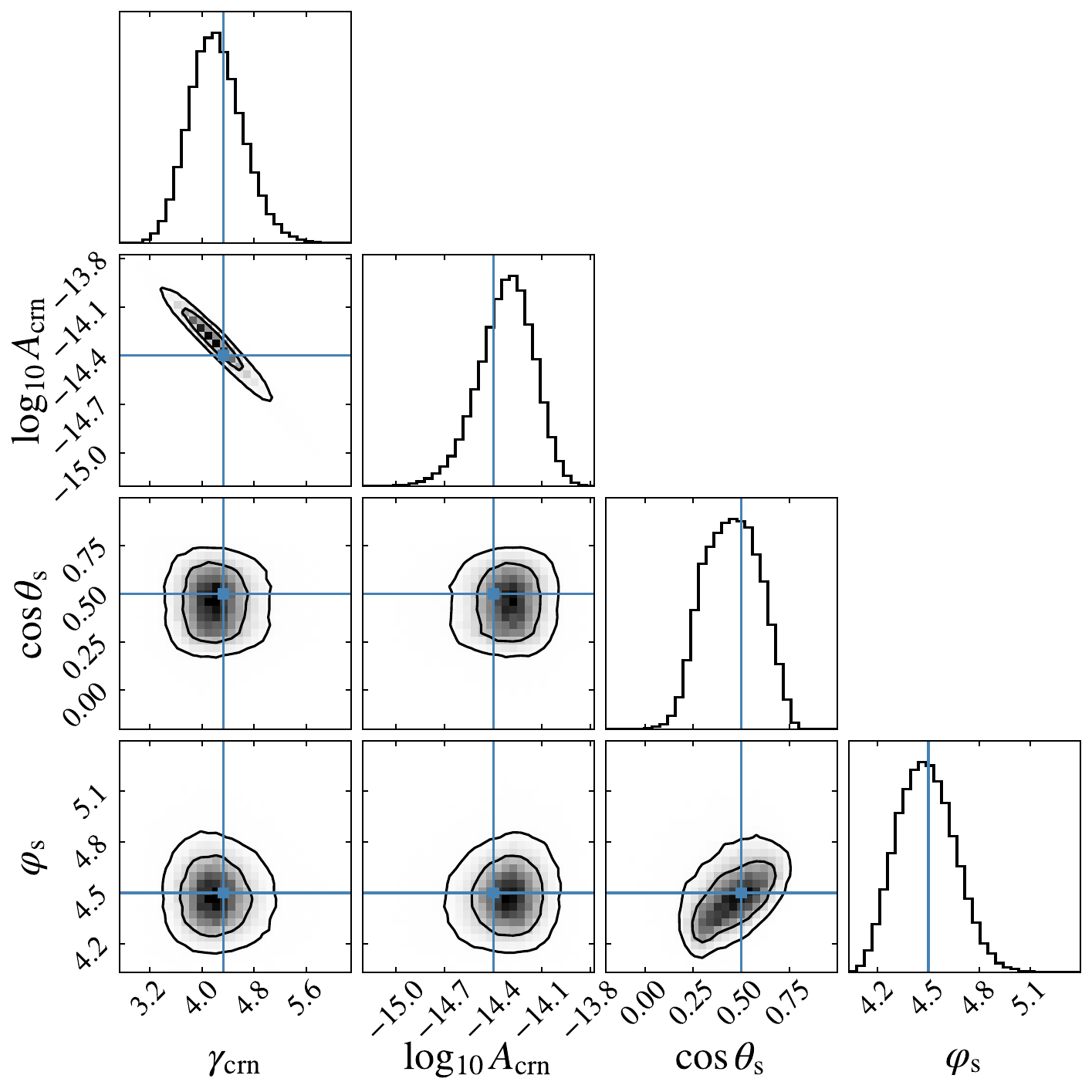}
\caption{\label{fig:cw_corner}Corner plot of $\gamma_{\rm crn}, \log_{10}A_{\rm crn}, \cos\theta_{\rm s}$ and $\varphi_{\rm s}$ in the waveform-agnostic model for dataset SINE. The blue lines show the injected parameter values. All parameters are accurately captured. }
\end{figure}

\begin{figure}[h!]
\centering
\includegraphics[scale=0.35]{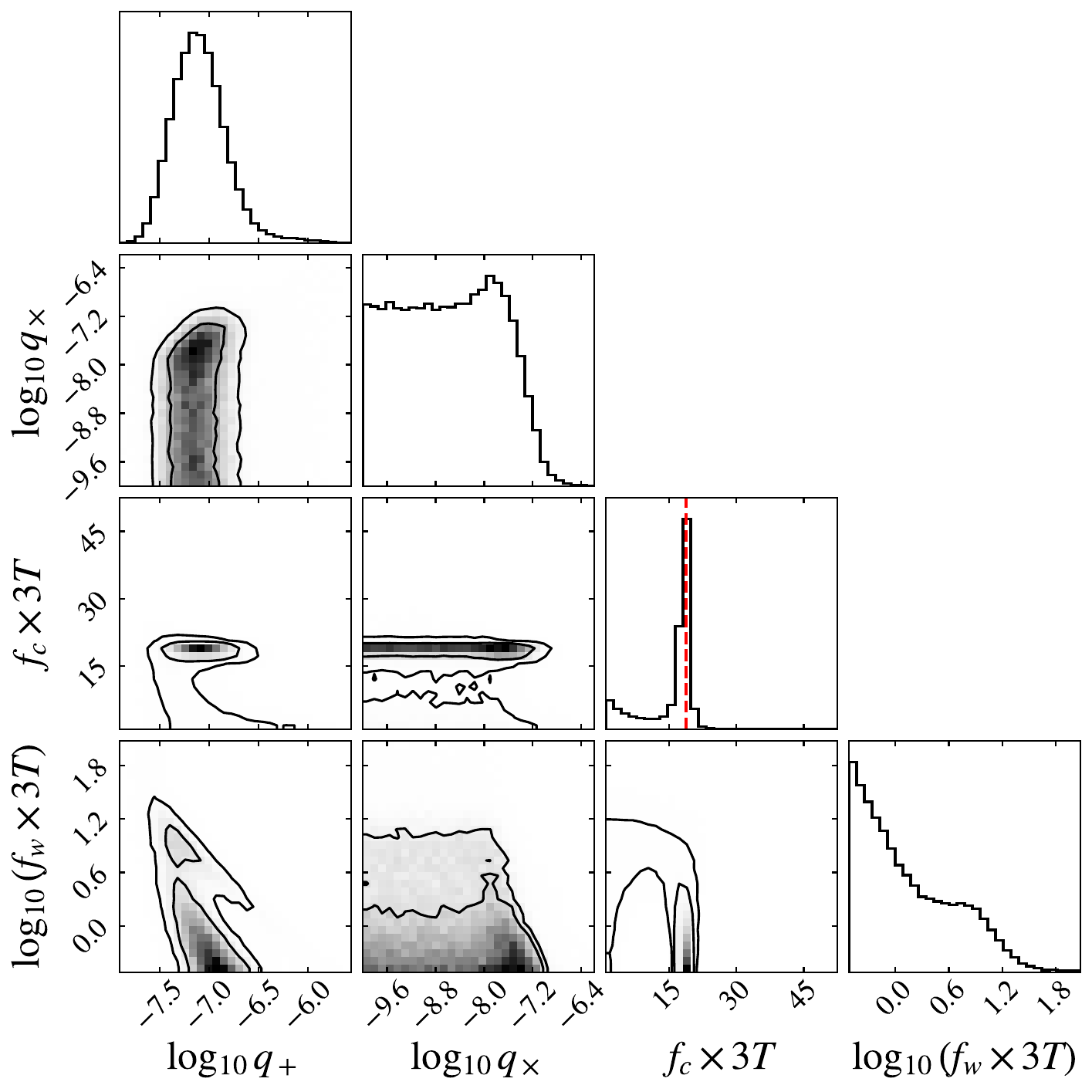}
\caption{\label{fig:cw_q+qcross} Corner plot of $\log_{10}q_+, \log_{10}q_\times,f_c$ and $\log_{10}f_w$ for dataset SINE. A strong plus polarization mode is picked up, and a peak around the injected sinusoidal frequency $f_{\rm s}\sim 20/3T$ (red dashed line) is captured by the posterior of $f_c$.}
\end{figure}

Fig. \ref{fig:cw_corner} shows the posteriors of the SGWB parameters $\gamma_{\rm crn}$ and $ \log_{10}A_{\rm crn}$, together with the source sky location parameters $\cos\theta_{\rm s}$ and $\varphi_{\rm s}$. The injected source location is correctly recovered, with the posterior peaking near the true values. Fig. \ref{fig:cw_q+qcross} shows the posteriors of the remaining four parameters in the waveform-agnostic model. As in the previous subsection, the plateau in $\log_{10}q_\times$ indicates a non-detection, which is consistent with the injection since the signal contains no cross polarization ($\psi_{\rm s}=0$). In contrast, the posterior of $\log_{10}q_+$ indicates a significant detection. The injected frequency of the sinusoidal signal is $f_{\rm s} = 2\times10^{-8} ~\text{Hz}$. In Fig. \ref{fig:cw_q+qcross}, we can see that the posterior of $f_c$ exhibits a clear peak located at $10^{1.3}/3T\approx2\times10^{-8}~\text{Hz}$, in excellent agreement with the injected value. Furthermore, the posterior of $f_w$ shows a preference for low frequencies, which is also consistent with the injection, since the sinusoidal signal is dominated by a single frequency. Overall, these results are fully consistent with the properties of the injected signal.

\begin{figure}[h!]
\centering
\includegraphics[scale=0.35]{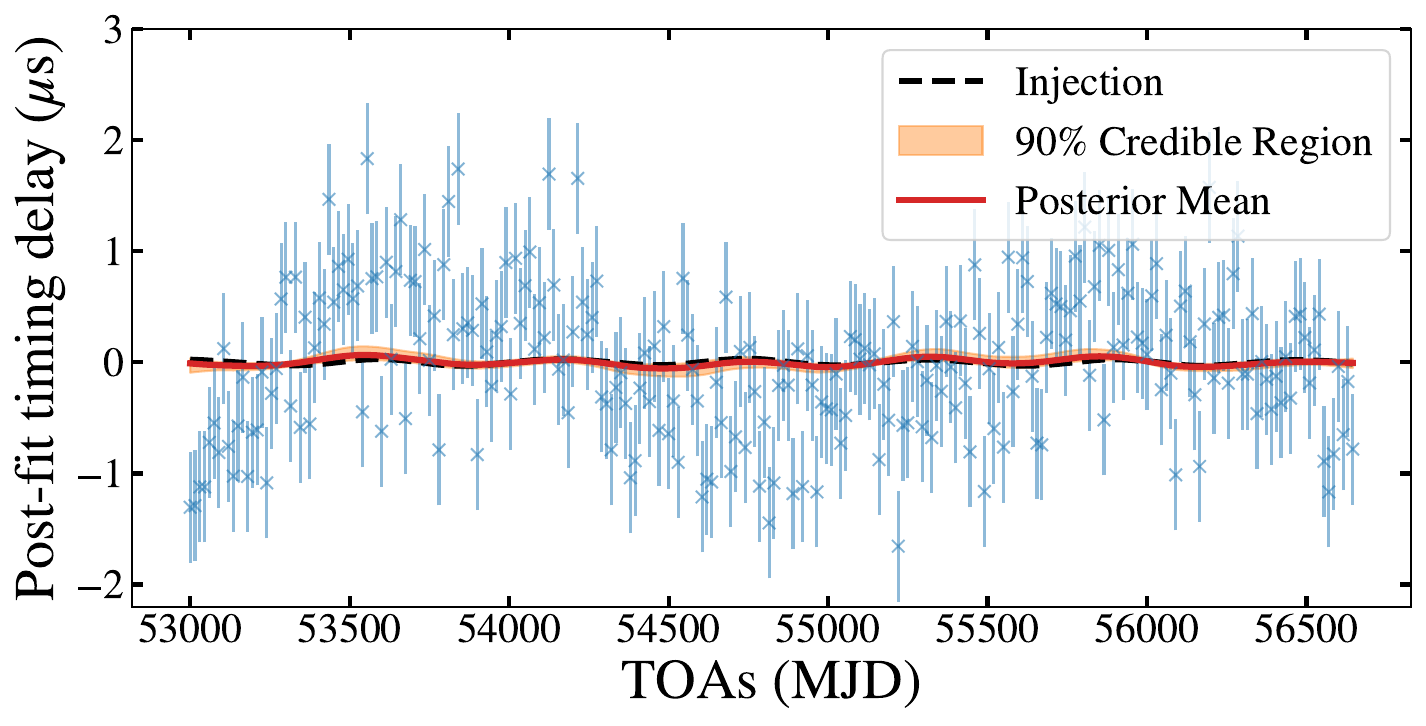}
\includegraphics[scale=0.35]{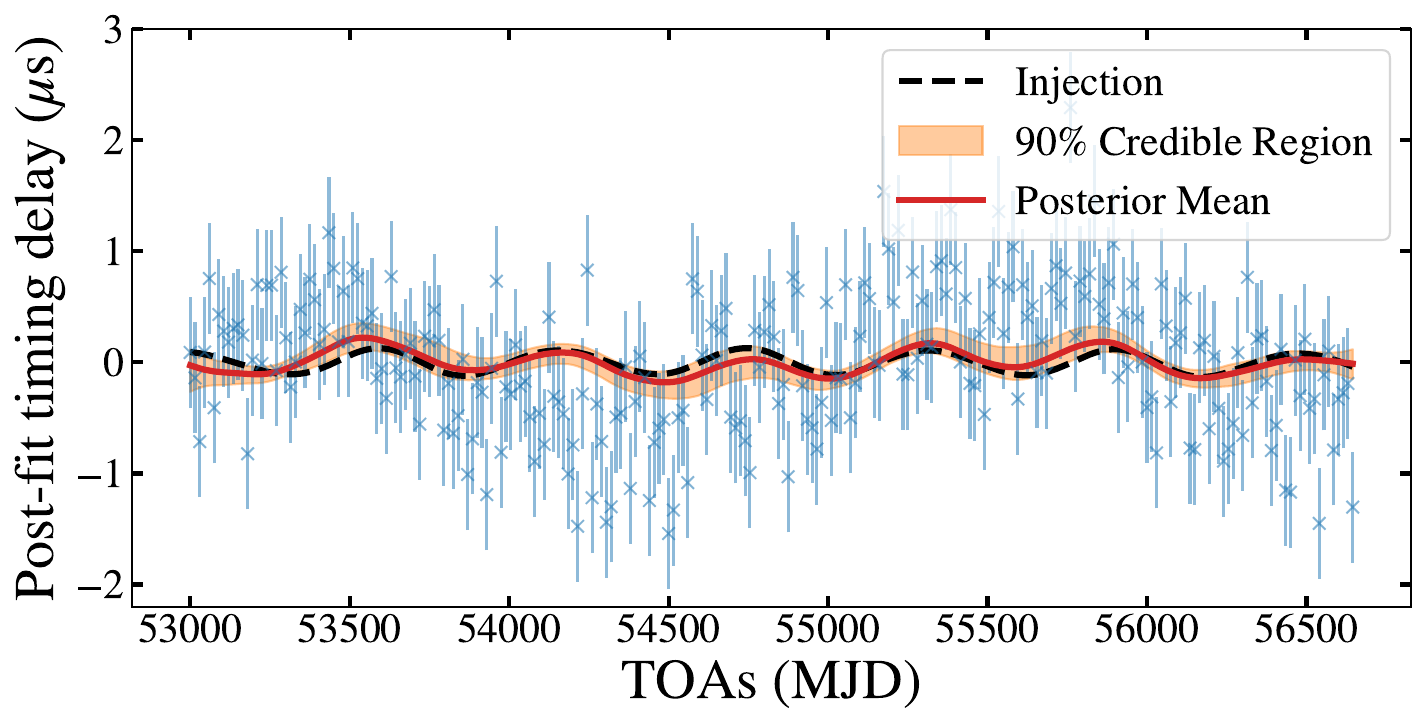}
\includegraphics[scale=0.35]{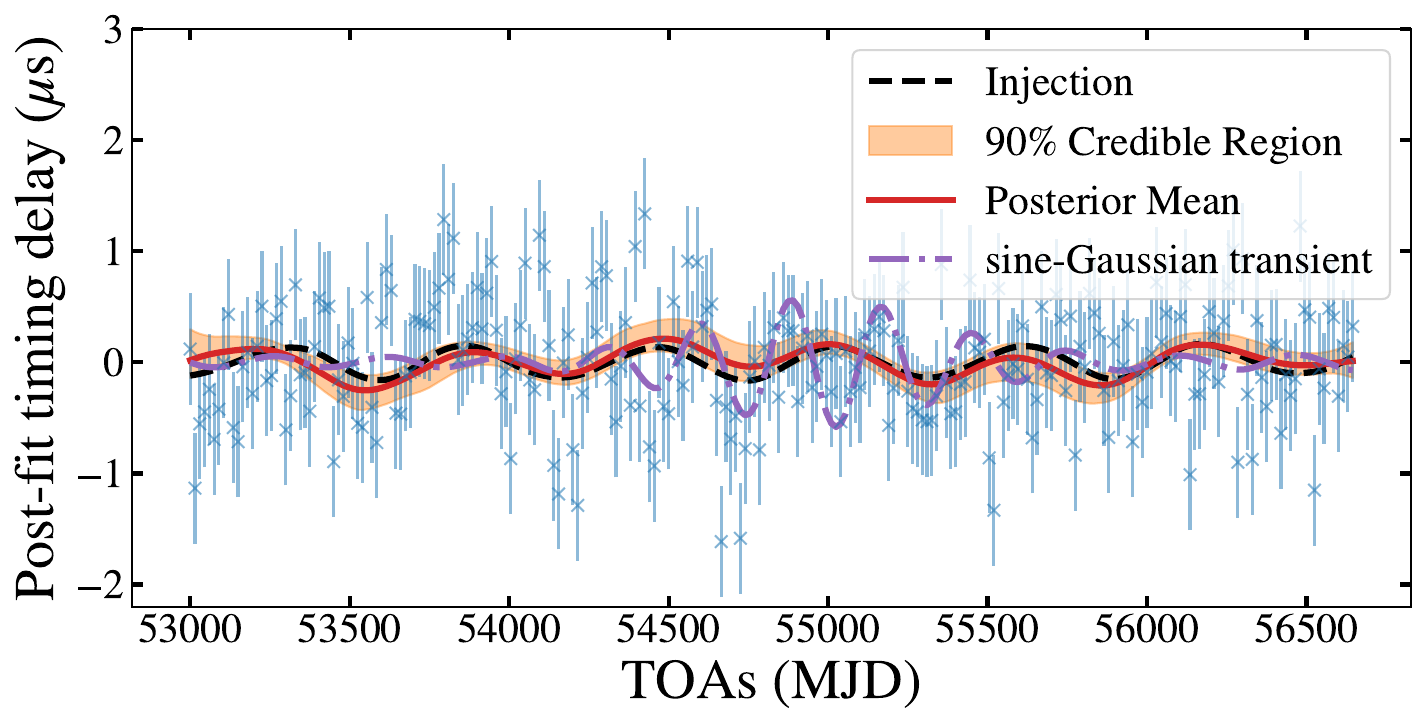}
\caption{\label{fig:cw_reconstruction}Reconstruction of the signal in three pulsars (Pulsars 0, 2 and 8 from the top) in the waveform-agnostic model for dataset SINE. The blue error bars represent the timing residuals with measurement uncertainties.}
\end{figure}

Signal reconstruction in three pulsars (Pulsars 0, 2, and 8) is shown in Fig. \ref{fig:cw_reconstruction}. Although the injected signals are mostly buried by white noise, the reconstructed signals (red) almost overlap with the injections (black). The sine-Gaussian noise transient added to Pulsar 8 (purple) is clearly distinguished from the $90\%$ credible band.

\begin{figure}[h!]
\centering
\includegraphics[scale=0.4]{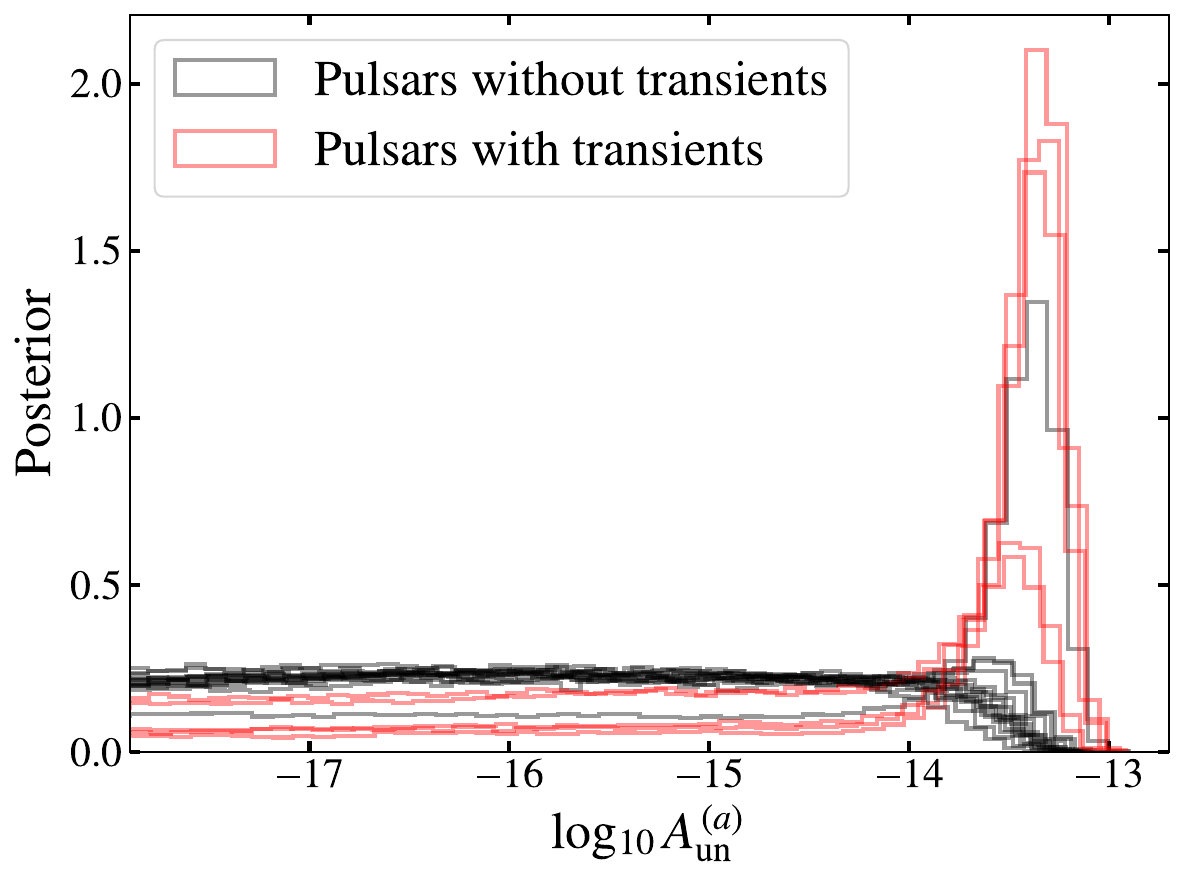}

\caption{\label{fig:cw_A}Posteriors of $\log_{10}A^{(a)}_{\rm un}$ for the twenty pulsars in the waveform-agnostic model for dataset SINE. Sine-Gaussian transients are injected into five pulsars.}
\end{figure}

The posteriors for the amplitudes of the unmodeled noise $\log_{10}A^{(a)}_{\rm un}$ are shown in Fig. \ref{fig:cw_A}. The red histograms correspond to the five pulsars containing noise transients. There are obvious peaks indicating the presence of excess power beyond what is expected from white noise or the SGWB. In addition to these five pulsars, a peak also appears in the posterior of $\log_{10}A^{(2)}_{\rm un}$. However, in all cases, the corresponding Savage–Dickey Bayes factors are too small to support detections of transients.

This experiment demonstrates that our waveform-agnostic model can successfully recover a narrow signal spectrum. Even in the presence of noise transients affecting a subset of pulsars (5 out of 20) with amplitudes comparable to the white noise level ($\sim0.5~\mu\text{s}$), the model accurately reconstructs the localized sinusoidal signal, despite its relatively small amplitude ($\sim 0.16~\mu\text{s}$).

\subsection{Comparison with waveform-template search}

In the previous two examples, we show how signals with very different frequency content can be faithfully recovered by our flexible model. A natural trade-off of this versatility is a reduction in accuracy: because the waveform-agnostic model blindly explores a large parameter space in search of an unknown signal, it can never outperform the waveform-template search when the true signal is well described by the template. To evaluate the sensitivity of our framework, we compare our model with a waveform-template search.

We inject a short-duration burst signal on top of the SGWB. Following Ref. \cite{Yonemaru:2020bmr}\footnote{Ref. \cite{Yonemaru:2020bmr} aimed to search for a burst signal emitted by the cusp of a cosmic string loop \cite{Damour:2001bk}. However, the waveform contribution from the rest of the loop was neglected.}, we consider a GW strain that scales as $\sim|t|^{1/3}$, so the induced residuals scale as $\sim|t|^{4/3}$:
\begin{align}
r^{(a)}(t)=&~ \left[F^{(a)}_+(\theta_{\rm b}, \varphi_{\rm b})\cos\psi_{\rm b}+F^{(a)}_{\times}(\theta_{\rm b}, \varphi_{\rm b})\sin\psi_{\rm b}\right]\notag\\&~\times\! A_{\rm b}\!\begin{cases}
0, & \!\!\!\tilde{t}<-\frac{1}{2}\\
-3\left(\frac{1}{2}\right)^{1/3}\left|\tilde{t}\right|^{4/3}-4\tilde{t}-\frac{1}{2}, &\!\!\! -\frac{1}{2}\le \tilde{t}<0\\
3\left(\frac{1}{2}\right)^{1/3}\left|\tilde{t}\right|^{4/3}-4\tilde{t}-\frac{1}{2}, &\!\!\! 0\le \tilde{t}<\frac{1}{2}\\
-1, &\!\!\! \tilde{t}\ge \frac{1}{2}
\end{cases}
\end{align}
where $\psi_{\rm b}$ is the polarization angle, $A_{\rm b}$ is the signal amplitude, and $\tilde{t}\equiv (t-t_{\rm b})/\sigma_{\rm b}$, with $\sigma_{\rm b}$ characterizing the signal duration and $t_{\rm b}$ denoting the time when the burst reaches its peak. We choose $\sigma_{\rm b}$ to be much smaller than the PTA observation span so that the residuals take the form of a step-like function (similar to Fig. \ref{fig:four_panel}(c)). We assume that the burst signal is linearly polarized and that the injected signal contains only the plus polarization, corresponding to $\psi_{\rm b}=0$. The signal amplitude is chosen such that the signal is barely detectable with our model. The injected parameter values are listed in Table \ref{tab:injected}.

We analyze the dataset using three models:
\begin{enumerate}
    \item[(i)] SGWB-only, with two parameters: $A_{\rm crn}$ and $\gamma_{\rm crn}$. The priors are listed in Table \ref{tab:injected}.
\item[(ii)] the waveform-template model, with eight parameters: $A_{\rm crn}, \gamma_{\rm crn}, \theta_{\rm b}, \varphi_{\rm b}, A_{\rm b}, \sigma_{\rm b}, t_{\rm b}$ and $\psi_{\rm b}$; this is the model that generates the dataset. The priors are listed in Table \ref{tab:injected}.
\item[(iii)] the waveform-agnostic model, with twenty eight parameters: $A_{\rm crn}, \gamma_{\rm crn}, A_{\rm un}^{(a)}, \theta_{\rm b}, \varphi_{\rm b}, q_+, q_\times, f_w$ and $f_c$, where $a= 1,2,\ldots, 20$. The priors are listed in Tables \ref{tab:injected} and \ref{tab:priors}.
\end{enumerate}
Sampling in models (i) and (ii) is performed using the nested-sampling software \texttt{Nestle} \cite{skilling2004nested,sivia2006data,Shaw:2007jj,Mukherjee:2005wg,Feroz:2008xx}, which provides both posterior samples and Bayesian evidences, allowing us to compute the Bayes factor between the two models.

\begin{figure}[h!]
\centering
\includegraphics[scale=0.35]{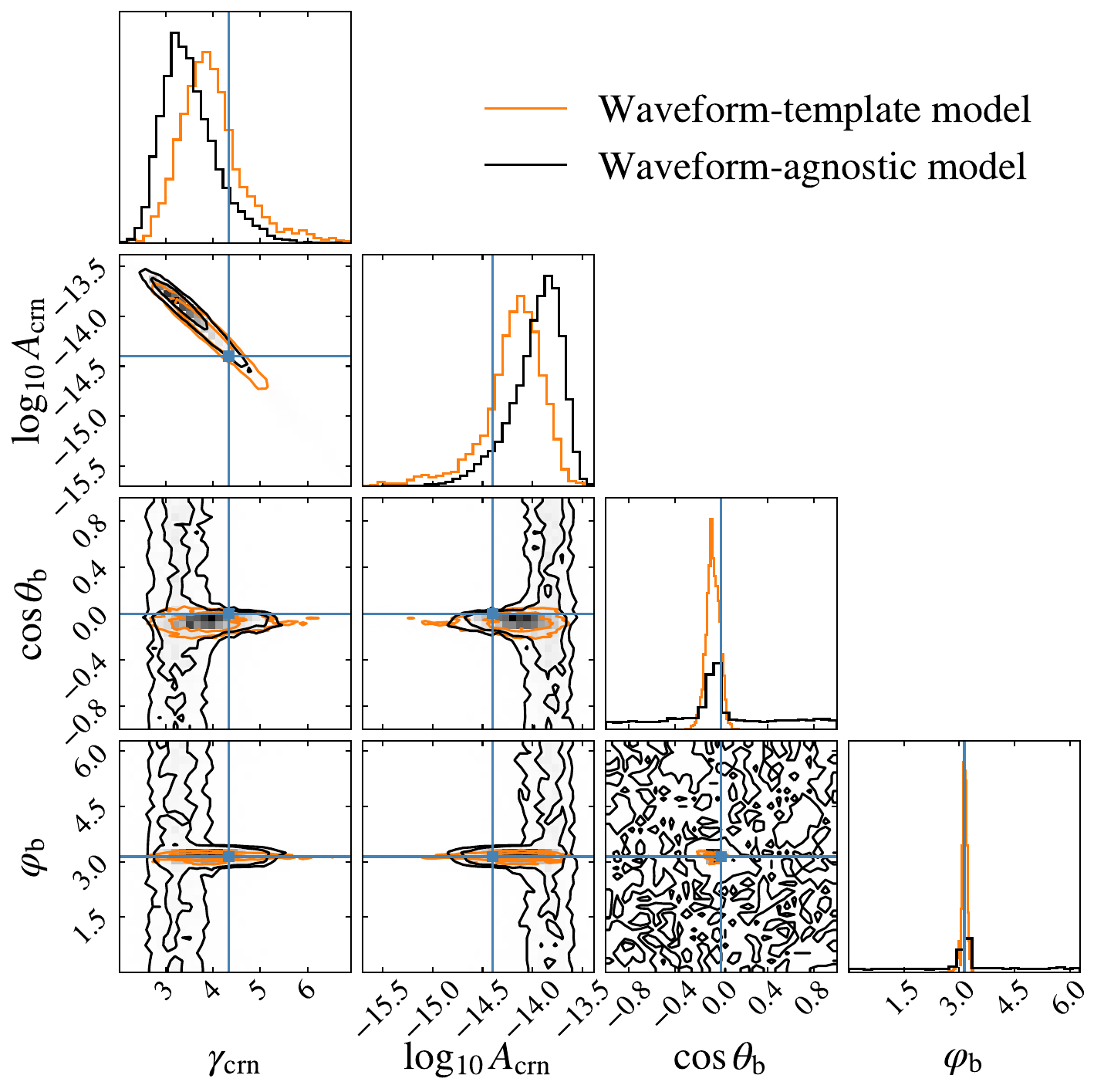}

\caption{\label{fig:burst_corner}Corner plot of $\gamma_{\rm crn}, \log_{10}A_{\rm crn}, \cos\theta_{\rm b}$ and $\varphi_{\rm b}$ in the waveform-template model (orange) and the waveform-agnostic model (black) for dataset BURST. The blue lines show the injected parameter values. The template model accurately localizes the source, whereas the agnostic model only marginally captures the true sky location, with a substantial fraction of samples scattering over the sky map.}
\end{figure}

Fig. \ref{fig:burst_corner} shows the corner plot of $\gamma_{\rm crn}, \log_{10}A_{\rm crn}, \cos\theta_{\rm b}$ and $\varphi_{\rm b}$ for both the waveform-template and waveform-agnostic models. Unsurprisingly, the waveform-template model accurately recovers all injected features, including a well-constrained source localization. The Bayes factor between the waveform-template model and the SGWB-only model is $\sim 50$, indicating very strong evidence for the presence of a burst signal.

In contrast, the waveform-agnostic model gives only a marginal detection of the signal. While the posterior of $(\theta_{\rm b}, \varphi_{\rm b})$ peaks at the true sky location, the samples are broadly distributed over the map. The SGWB amplitude $\log_{10}A_{\rm crn}$ is slightly overestimated. Notably, samples away from the injected sky location in the $\log_{10}A_{\rm crn}\text{-}\cos\theta_{\rm b}$ and $\log_{10}A_{\rm crn}\text{-}\varphi_{\rm b}$ plots mostly contribute to the peak of $\log_{10}A_{\rm crn}$'s posterior, suggesting that the SGWB absorbs part of the burst power.

\begin{figure}[h!]
\centering
\includegraphics[scale=0.35]{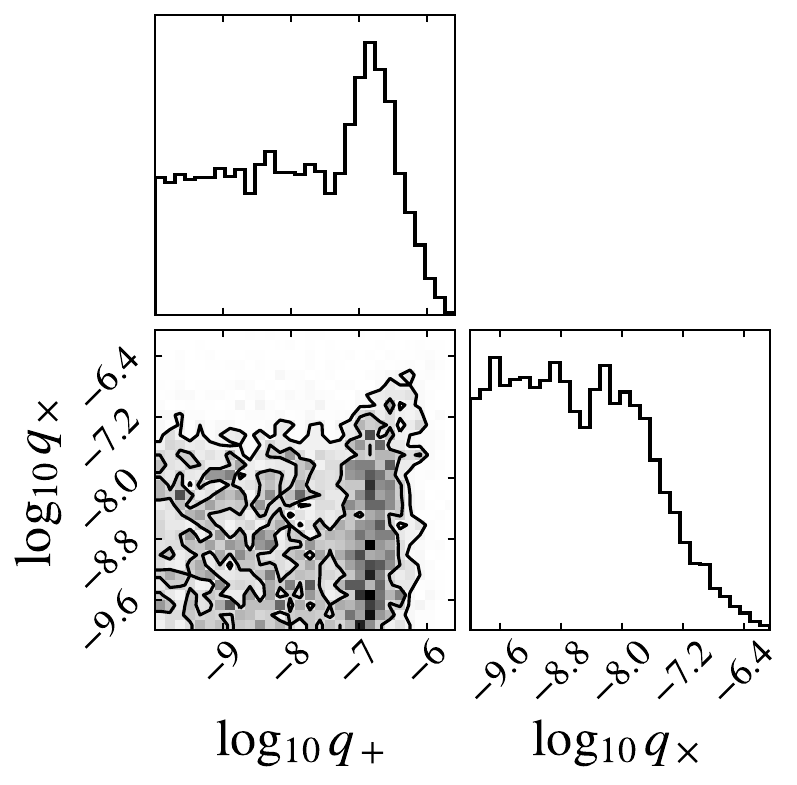}

\caption{\label{fig:q+qcross}Corner plot of $\log_{10}q_+$ and $\log_{10}q_\times$ in the waveform-agnostic model for dataset BURST. A weak plus polarization mode is picked up.}
\end{figure}

Fig. \ref{fig:q+qcross} shows the posteriors for the signal's spectral amplitudes. The posterior of $\log_{10}q_\times$ clearly indicates a non-detection, consistent with the injected signal, which contains no cross polarization. Although the posterior of $\log_{10}q_+$ has a peak, its heavy tail towards the left suggests that the detection is not statistically significant. By the Savage-Dickey ratio, the Bayes factor relative to the SGWB-only model is only $\sim 1$.

\begin{figure}[h!]
\centering
\includegraphics[scale=0.35]{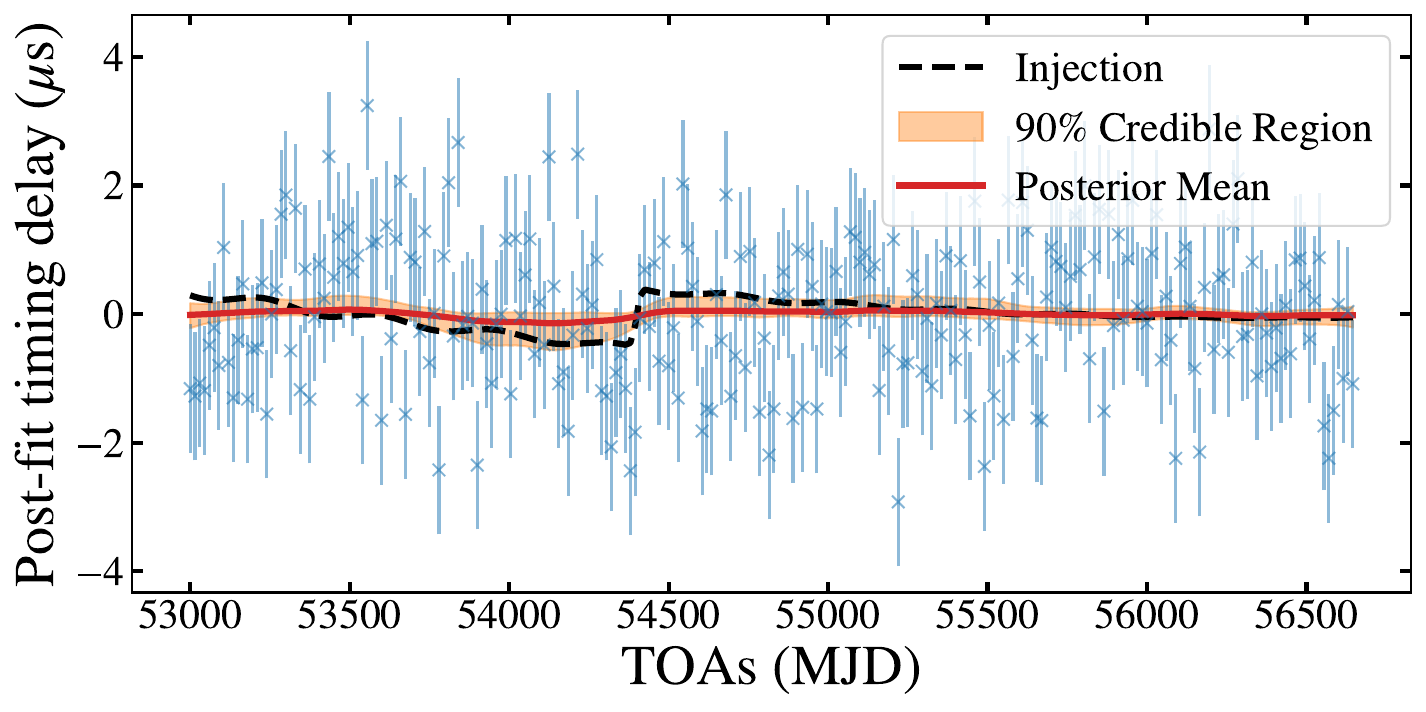}
\includegraphics[scale=0.35]{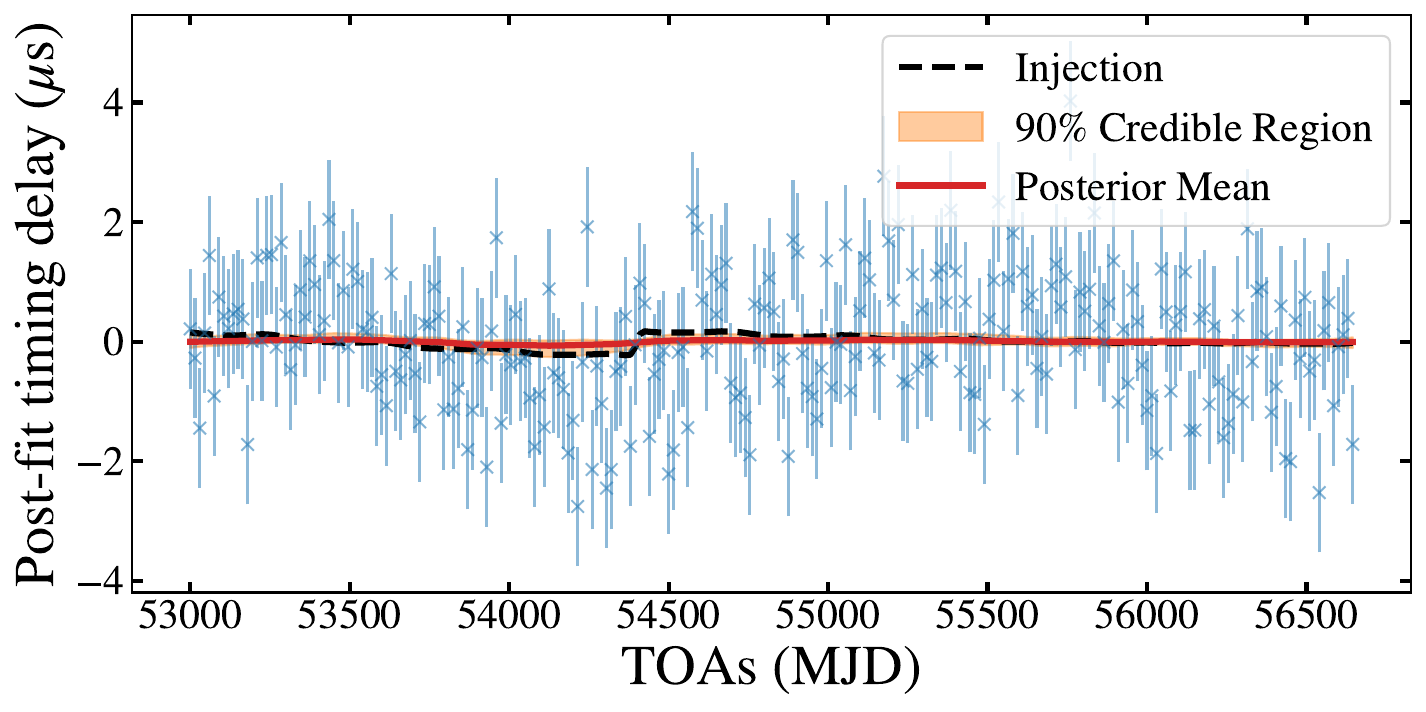}
\includegraphics[scale=0.35]{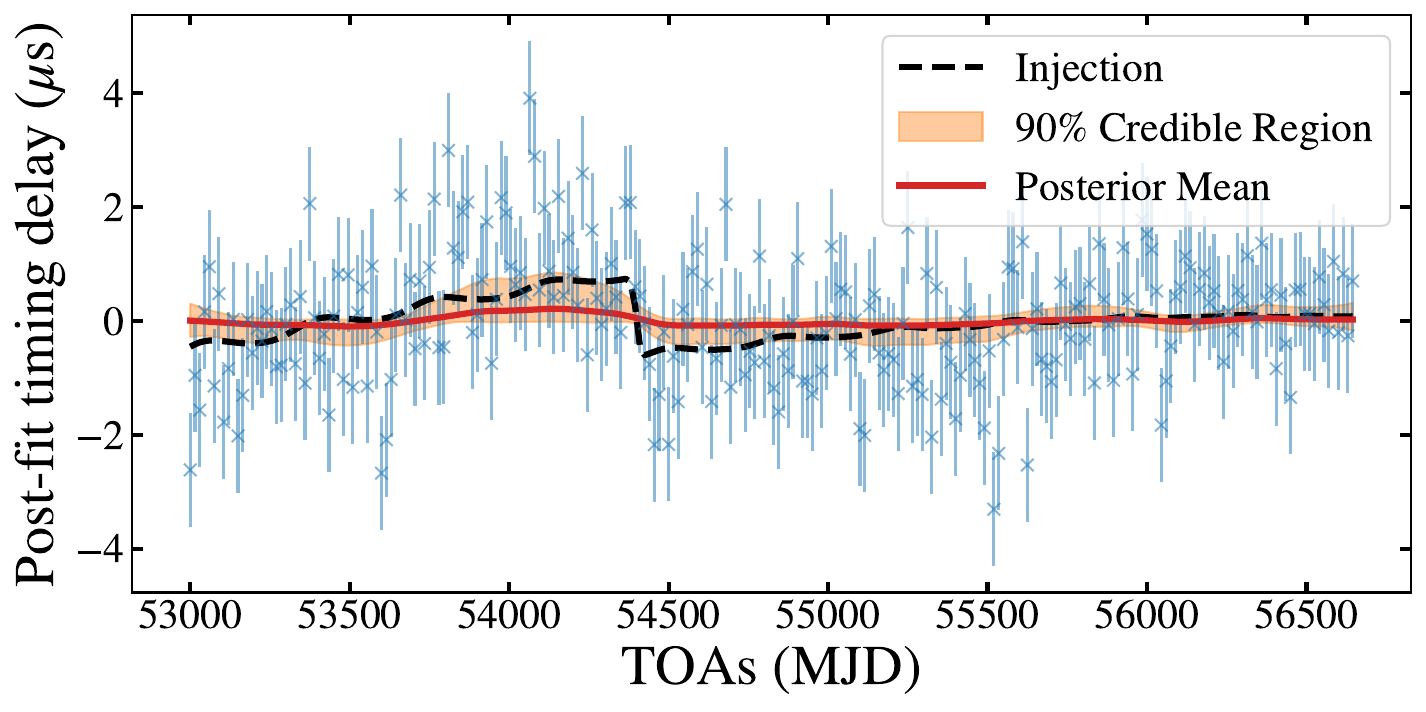}
\caption{\label{fig:burst_reconstruction}Reconstruction of the signal in three pulsars (Pulsars 0, 2 and 9 from the top) in the waveform-agnostic model for dataset BURST. The blue error bars represent the timing residuals with measurement uncertainties.}
\end{figure}

The signal reconstruction for three pulsars in the waveform-agnostic model is shown in Fig. \ref{fig:burst_reconstruction}. Because the injected burst is intentionally chosen to be barely detectable by the waveform-agnostic model, the reconstructed signals (red) do not match the injected ones (black). Nevertheless, they show some features at the time when the burst occurs, allowing the model to correctly ``guess'' the sky location by adjusting the antenna pattern response.

The different outcomes of the waveform-template and waveform-agnostic models are not unexpected. The template model already knows what it is looking for, while the agnostic model has to explore a wider range of possibilities without any prior knowledge. Despite this disadvantage, our method still successfully points to the correct sky location. This shows that, while template searches are more effective for well-understood and anticipated signals, the agnostic approach provides a valuable complementary tool for unexpected sources. Even if the signal's shape cannot be accurately reconstructed, it can still provide a hint of the source's sky location, allowing us to perform subsequent targeted searches.

\subsection{Potential false alarm from noise transients \label{sine-Gaussian}}

In this section, we consider two datasets, D1 and D2, designed to examine a potential limitation of the waveform-agnostic model. Unlike the previous datasets, no coherent GW signal is injected. Instead, we add sine-Gaussian transients to ten pulsars, while the remaining pulsars contain only white noise and the SGWB. In dataset D1, the injected transients are identical across the ten affected pulsars, with parameter values listed in Table~\ref{tab:injected}. As in dataset SINE, the amplitudes are comparable to the white noise level.

\begin{figure*}
\centering
\includegraphics[scale=0.35]{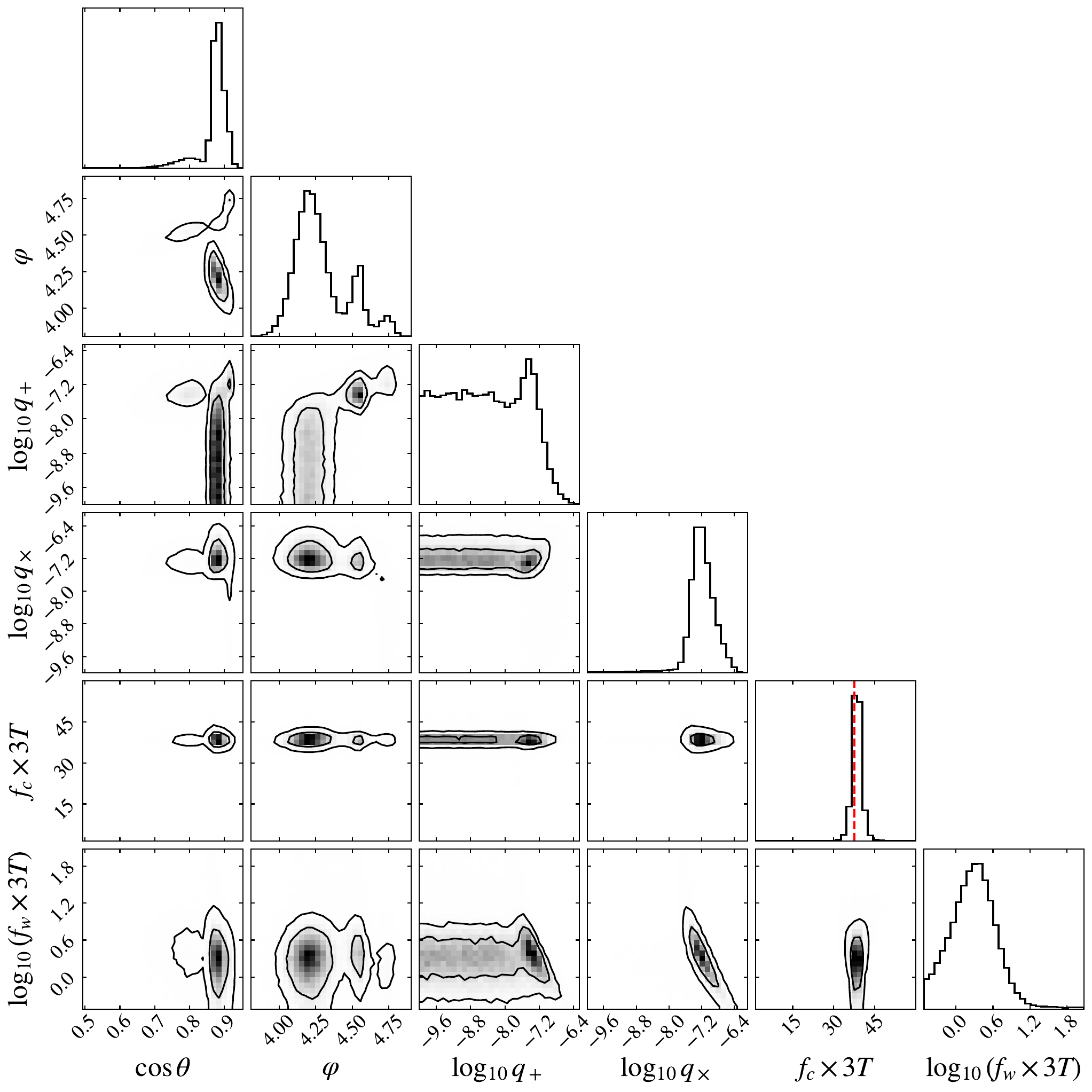}

\caption{\label{fig:noise_corner}Corner plot of the parameters in the waveform-agnostic model for dataset D1. Although the dataset does not contain a coherent signal, a strong localized signal is detected. The central frequency of the Lorentzian prior, $f_c$, identifies the frequency of the sine-Gaussian noise injected into ten pulsars (red dashed line).}
\end{figure*}

\begin{figure}
\centering
\includegraphics[scale=0.35]{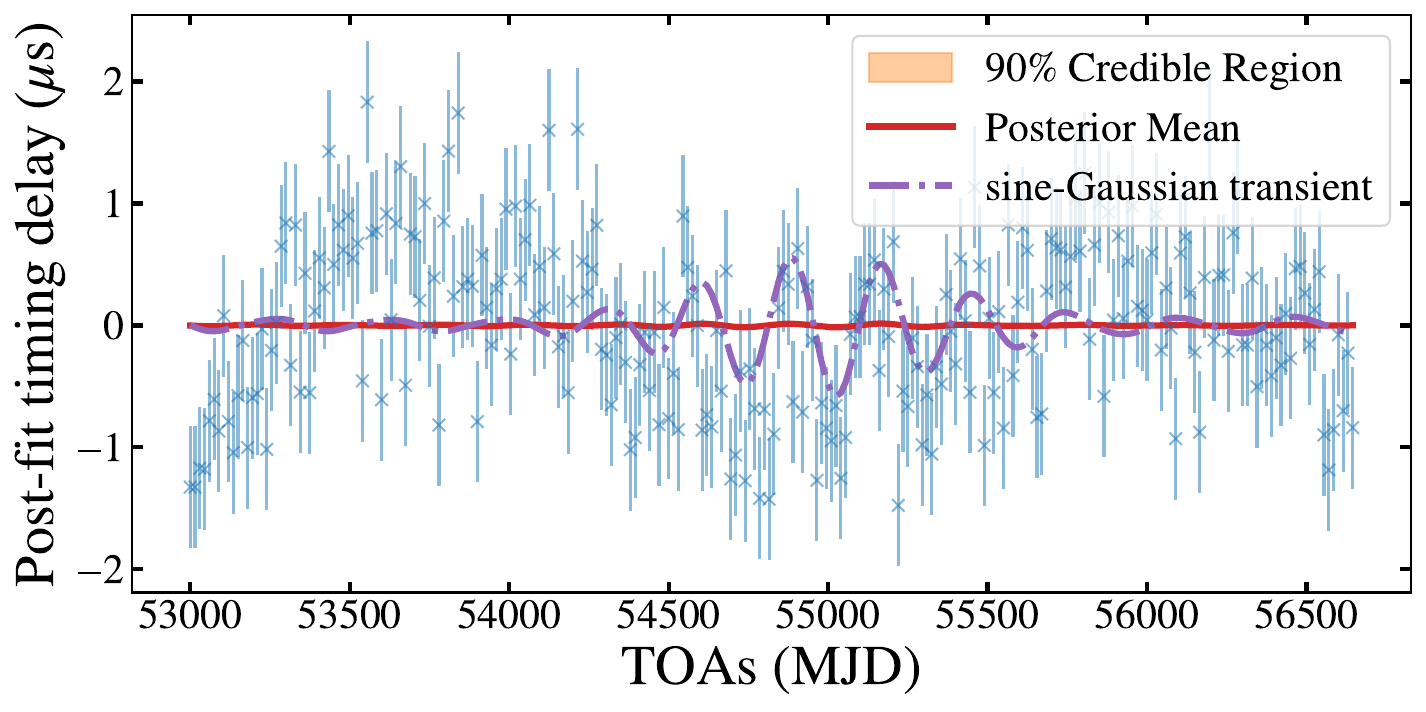}
\includegraphics[scale=0.35]{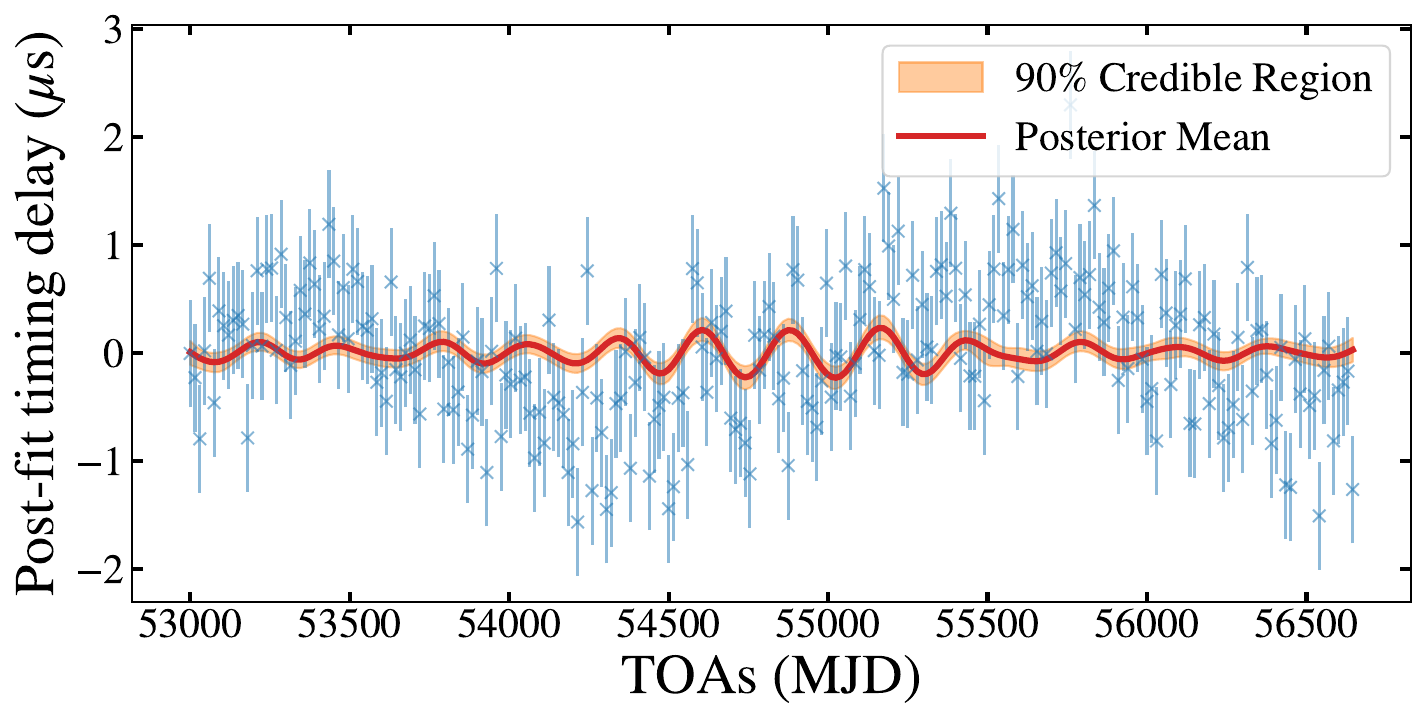}
\includegraphics[scale=0.35]{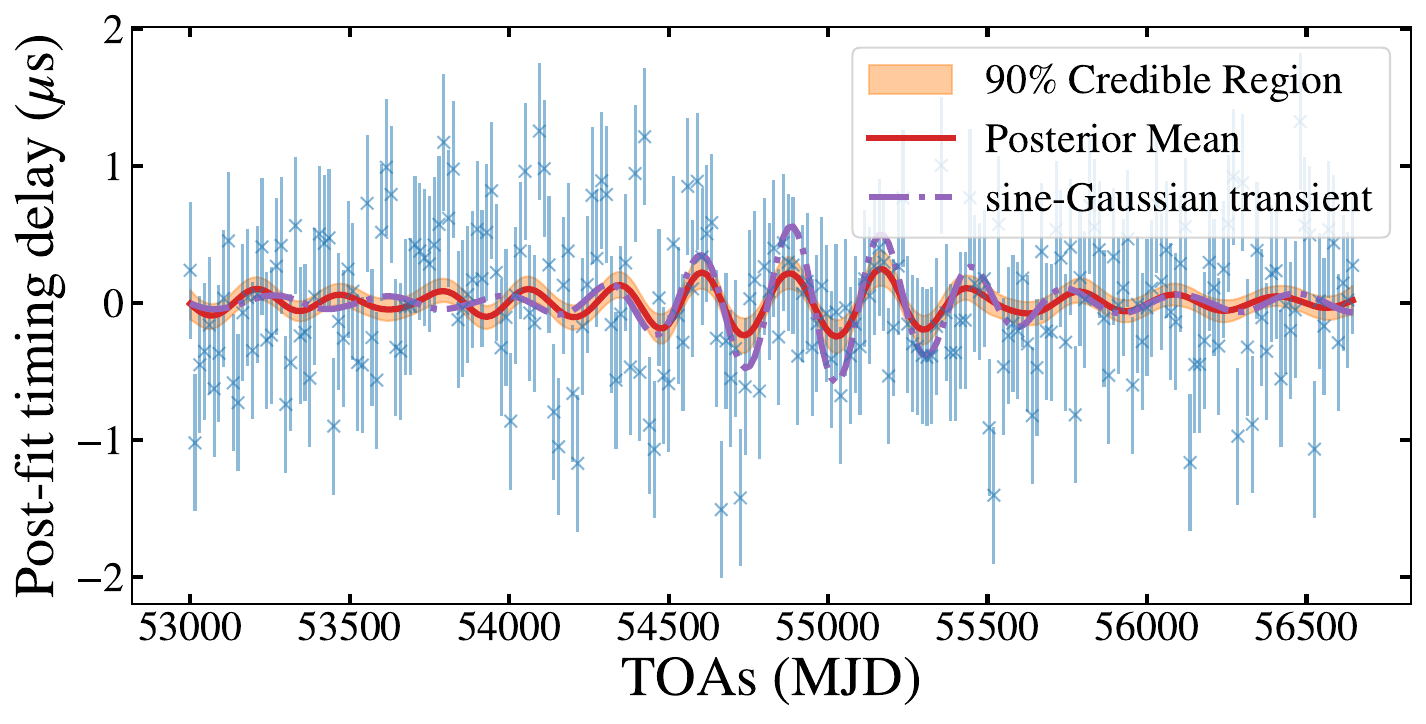}
\caption{\label{fig:sg_reconstruction}Reconstruction of the signal in three pulsars (Pulsars 0, 2 and 8 from the top) in the waveform-agnostic model for dataset D1. The blue error bars represent the timing residuals with measurement uncertainties. No coherent signal is injected; the false positive signal tries to explain the injected sine-Gaussian noise in some pulsars (e.g., Pulsar 8 here).}
\end{figure}

Fig. \ref{fig:noise_corner} shows the posteriors of the parameters in the waveform-agnostic model. Despite the absence of any injected coherent signal, the model identifies a localized source, with posterior support concentrated near a specific sky position. While $\log_{10}q_+$'s posterior is consistent with non-detection, $\log_{10}q_\times$'s posterior has a significant peak. In addition, the posterior of $f_c$ picks up a dominant frequency $38/3T\sim 10^{-7.4} ~\text{Hz}$, which is precisely the sinusoidal frequency of the injected sine-Gaussian noise. All these demonstrate that, under sufficiently contrived conditions, pulsar noise can be misidentified as a localized GW signal. Fig. \ref{fig:sg_reconstruction} shows the reconstruction of the localized signal in Pulsars 0, 2 and 8. In Pulsar 0, the signal has a negligible impact, whereas in Pulsar 8, the reconstructed signal can closely match the injected transient.

\begin{figure}
\centering
\includegraphics[scale=0.4]{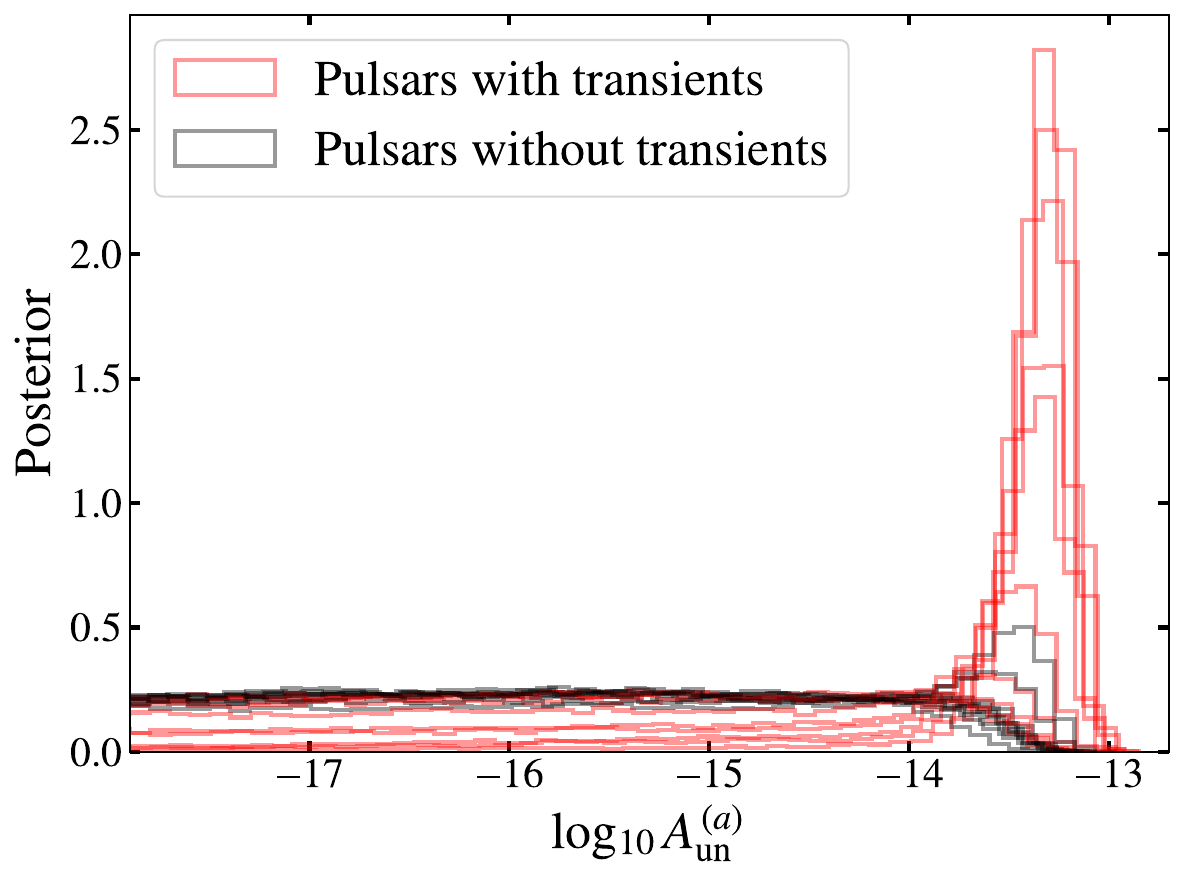}

\caption{\label{fig:sg_A}Posteriors of $\log_{10}A^{(a)}_{\rm un}$ for the twenty pulsars in the waveform-agnostic model for dataset D1. Identical sine-Gaussian transients are injected into ten pulsars.}
\end{figure}

Fig. \ref{fig:sg_A} shows the posteriors of $\log_{10}A_{\rm un}^{(a)}$ for all twenty pulsars. The red histograms correspond to the ten pulsars with noise transients. We can see that not all transients are captured by the flat spectra. Apparently, those undiscovered transients are absorbed by the spurious coherent signal identified by the model.

As pointed out in Ref. \cite{Becsy:2020utk}, this behavior reflects the inherent parsimony of Bayesian analysis, as the waveform-agnostic model trades an increase in model complexity against an improvement in likelihood. When a sufficiently large number of pulsars exhibit similar transient features, the model can find it statistically favorable to attribute them to a coherent signal with an appropriate sky location, rather than to, say, ten independent noise processes. 

It is important to emphasize that this dataset represents an intentionally designed worst-case scenario. In real data, noise transients are expected to vary significantly from pulsar to pulsar in both shape and amplitude. It is unlikely that ten pulsars would have nearly identical transient features. Consequently, the false detection observed here should be interpreted as a conservative limitation of the waveform-agnostic approach.

\begin{figure}
\centering
\includegraphics[scale=0.35]{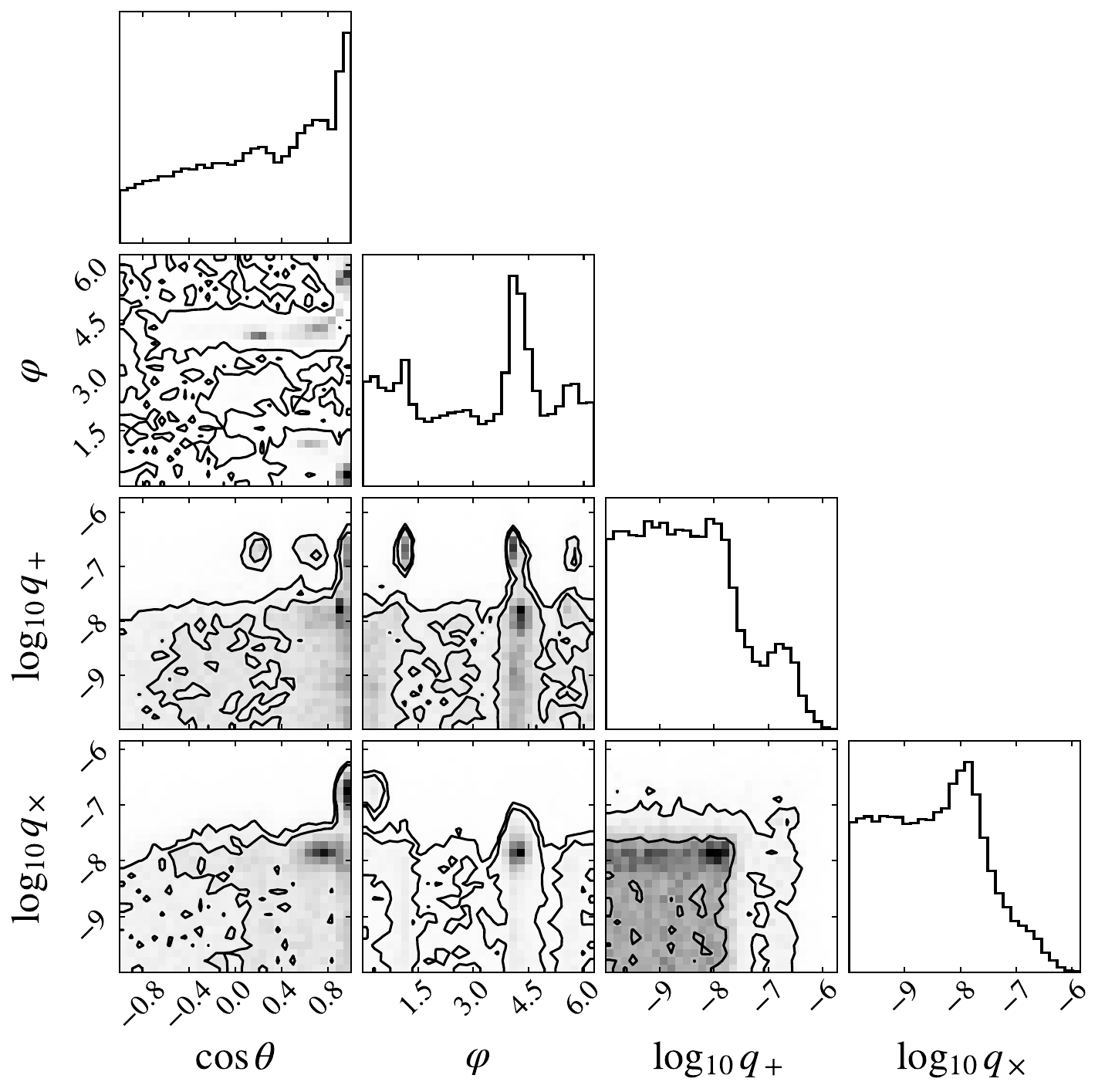}

\caption{\label{fig:noise2_corner}Corner plot of the parameters in the waveform-agnostic model for dataset D2. No coherent signal is picked up.}
\end{figure}

To test this, we construct a modified dataset, D2, in which the ten sine-Gaussian transients still occur at the same epoch, but differ in frequency, with values randomly drawn from a specified range (see Table \ref{tab:injected}). Fig. \ref{fig:noise2_corner} shows the posteriors of the parameters in the waveform-agnostic model, which are consistent with no-detection. Compared with Fig. \ref{fig:noise_corner}, the inference changes drastically. 

These tests illustrate a potential limitation of waveform-agnostic searches: when a sufficiently large subset of pulsars contains similar noise transients, the model may interpret these features as evidence of a coherent, localized signal. However, this is unlikely to happen with realistic PTA datasets. Moreover, the sine-Gaussian noise injected into the pulsars here has an amplitude comparable to the white noise level. In practice, noise of this kind would likely be (at least partially) absorbed in individual pulsar noise modeling through the jitter noise, which accounts for white noise correlated between TOAs at different frequencies within the same observing epoch. In other words, unmodeled noise in real data is expected to have amplitudes below the white noise level, which reduces the possibility of it being misidentified as a coherent signal.

\section{Conclusions and discussion}

We have presented a waveform-agnostic framework for searching for localized, deterministic gravitational wave signals in PTA data. Rather than assuming a specific waveform template, our approach models the signal-induced timing residuals using a flexible Fourier expansion with coefficients analytically marginalized. To capture the signal’s spectral structure, we impose an informative hyperprior on the autocovariance of the Fourier coefficients that has the shape of a Lorentzian function, which features a peak, a width, and heavy tails. Furthermore, by incorporating pulsar-specific flat spectra, our model effectively disentangles coherent gravitational wave signals from unmodeled, incoherent pulsar noise. This integrated Bayesian framework enables the simultaneous inference of a source’s sky location and its frequency content. Once a signal is identified, its time-domain shape can be reconstructed directly from the resulting posterior samples.

We evaluated the performance of the method using five mock datasets in several scenarios. We first considered two signals with very different spectra: a sinc function, whose Fourier transform is a broad rectangular profile, and a pure sinusoid, whose spectrum has support only at a single frequency. In both cases, the spectral structure was accurately captured by the Lorentzian hyperprior; the sky location of the source and the shape of the signal were also faithfully recovered. The third dataset contains a short-duration burst waveform, whose induced residuals look like a step function. This signal can be confidently detected with a waveform template. We found that the waveform-agnostic model was able to marginally recover the sky location, despite the signal being only weakly detected. These results demonstrate that the method can identify a wide range of coherent signals without relying on detailed waveform assumptions.

We also conducted noise-only tests that demonstrate an intrinsic limitation of our model. When a sufficiently large number of pulsars have nearly identical noise transients, the Bayesian model can incorrectly favor a single coherent signal with an appropriately chosen sky location. This occurs because the model can trade multiple independent noise processes for a single signal described by a small number of parameters. We emphasize that this scenario represents a worst-case scenario because, in realistic PTA observations, pulsar noise transients are expected to vary substantially across pulsars, both in shape and amplitude.

The results presented here suggest that waveform-agnostic analysis can serve as a valuable complement to conventional, template-based searches. They provide a flexible way of searching for unexpected or poorly modeled signals while retaining sensitivity to familiar sources such as continuous waves. When a candidate signal is identified, waveform-agnostic inference can guide a subsequent, more specialized analysis by providing an initial estimate of the sky location and characteristic frequencies.

The Lorentzian hyperprior assigned to the variances of Fourier coefficients is intended to capture the signal's frequency content. Of course, this choice is not unique. One may consider adding more features, such as skewness or inter-frequency correlations, to the covariance. Other versatile spectral models include spline interpolation and free spectrum. We chose a simple prior due to the relatively small parameter space it introduces. Furthermore, while the same spectral framework could be applied to unmodeled pulsar noise, we chose the flat spectrum in order to limit the computational cost.

Lastly, using a Fourier expansion to describe localized deterministic GW signals can be naturally extended to waveform-template-based models. In the conventional formulation, waveform-induced timing residuals $s$ enter the likelihood in Eq.~(\ref{eq:Lapprox}) through the replacement $r \to r - s$. When the waveform parameters are varied during Markov chain Monte Carlo sampling, quantities such as $\langle s|s\rangle$ must be evaluated at every step, which can be computationally expensive. In contrast, if we model $s$ using a Fourier series of the form $s=SH$ (see Eq. (\ref{FP})), then $\langle s|s\rangle=H^\top \langle S|S\rangle H$, where the coefficient vector $H$ has a dimension significantly smaller than that of $s$, and the matrix $\langle S|S\rangle$ can be pre-computed. This substantially reduces the cost of likelihood evaluations. For Bayesian inference, one simply assigns priors to the physical parameters in the waveform rather than to the Fourier coefficients, as done in the present work. For each draw of the waveform parameters, the corresponding Fourier coefficients are obtained via the Fourier decomposition of the waveform. This approach is fully general and can be applied to arbitrary waveform models, including those with pulsar terms. A recent application of this method to continuous waves from supermassive black hole binaries is presented in Ref.~\cite{Gundersen:2024qmq}.

\section*{Acknowledgments}
We thank Bjorn Larsen for insightful comments. HD and YL were supported by Yuri Levin’s Simons Investigator Grant PG012519.

\bibliography{main}

\appendix
\section{Examples of timing residuals induced by GW signals \label{append}}
It is instructive to illustrate how different shapes of GW signals affect the timing residuals. For simplicity, here we ignore polarization and the antenna pattern and consider a simple sine–Gaussian waveform:
\begin{equation}
    h(t) = A_{\rm sg}\sin\left[2\pi f_{\rm sg} (t - t_{\rm sg}) + \phi_{\rm sg}\right] e^{-\frac{(t - t_{\rm sg})^2}{2\sigma_{\rm sg}^2} }.\label{eq:sg}
\end{equation}
By varying the Gaussian width $\sigma_{\rm sg}$ and the oscillation frequency $f_{\rm sg}$ relative to the PTA observation span, this function can approximate several physically relevant signals. The induced timing residuals are proportional to the time integral of $h(t)$, and the post-fit residuals are obtained by removing the constant, linear, and quadratic components (due to the spin-down fit).

\subsection{Sine-Gaussian wave packet}
\begin{figure*}[t]
    \centering
    \subfloat[]{%
        \includegraphics[width=0.25\textwidth]{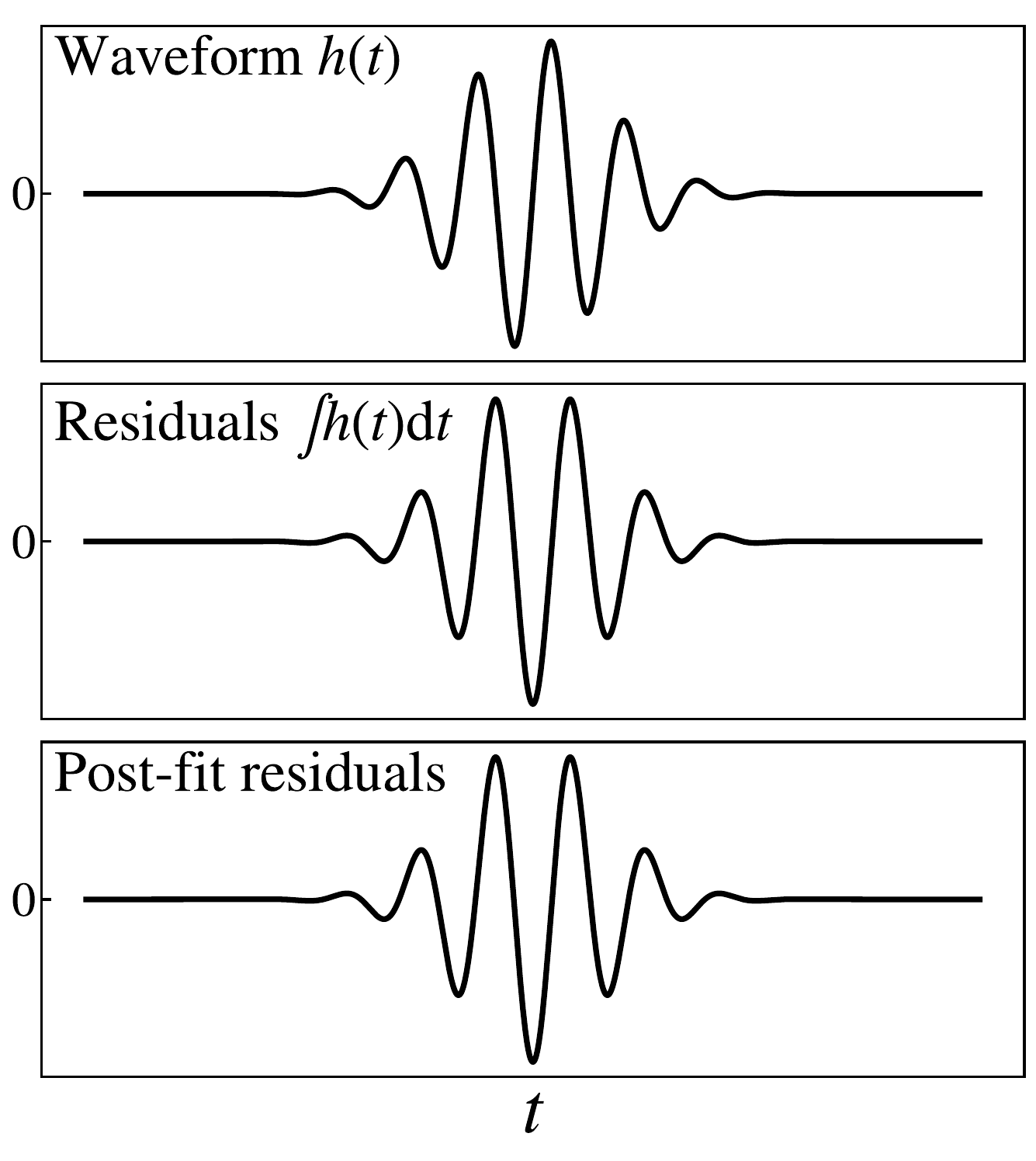}
    }
    \subfloat[]{%
        \includegraphics[width=0.25\textwidth]{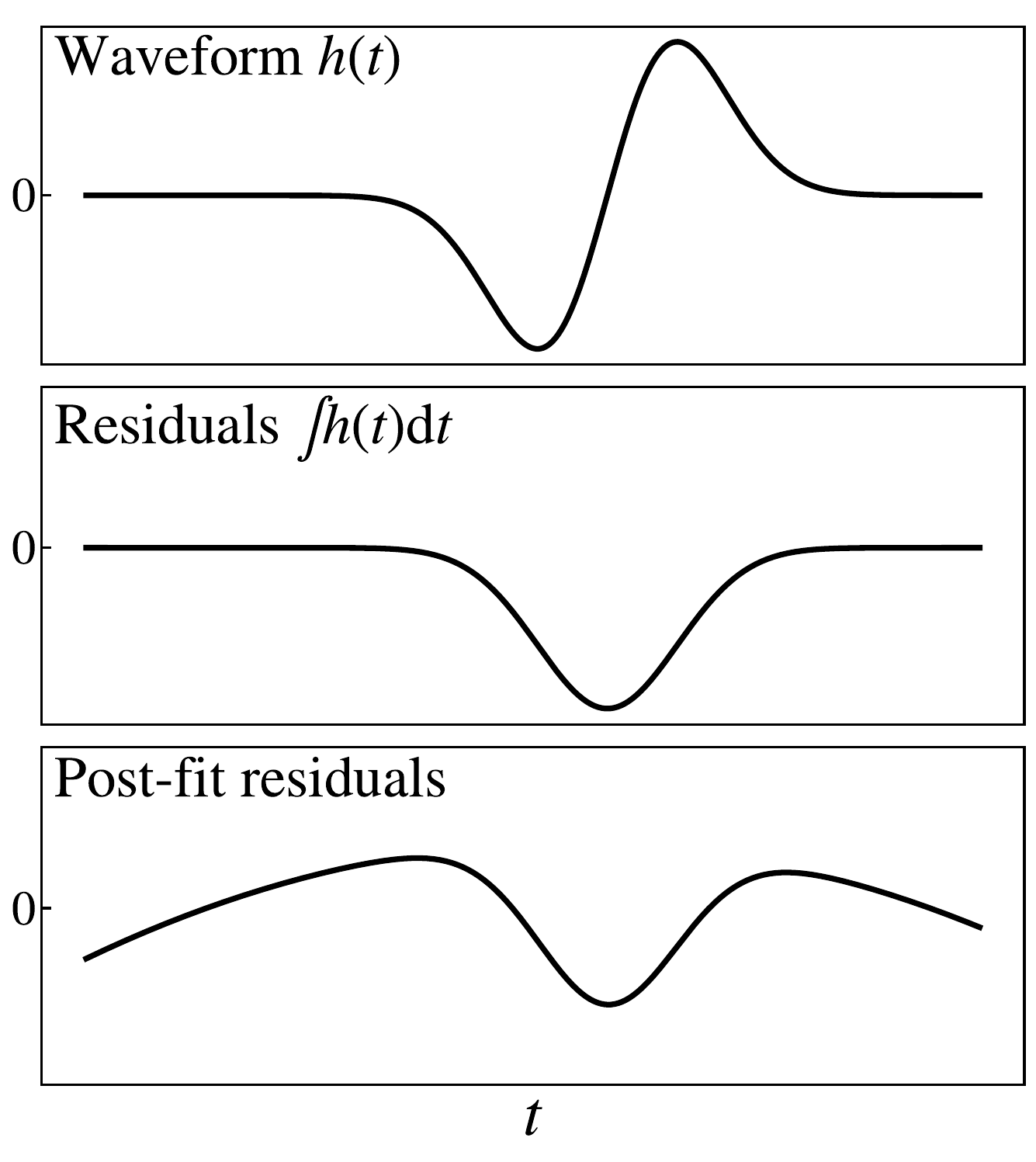}
    }
    \subfloat[]{%
        \includegraphics[width=0.25\textwidth]{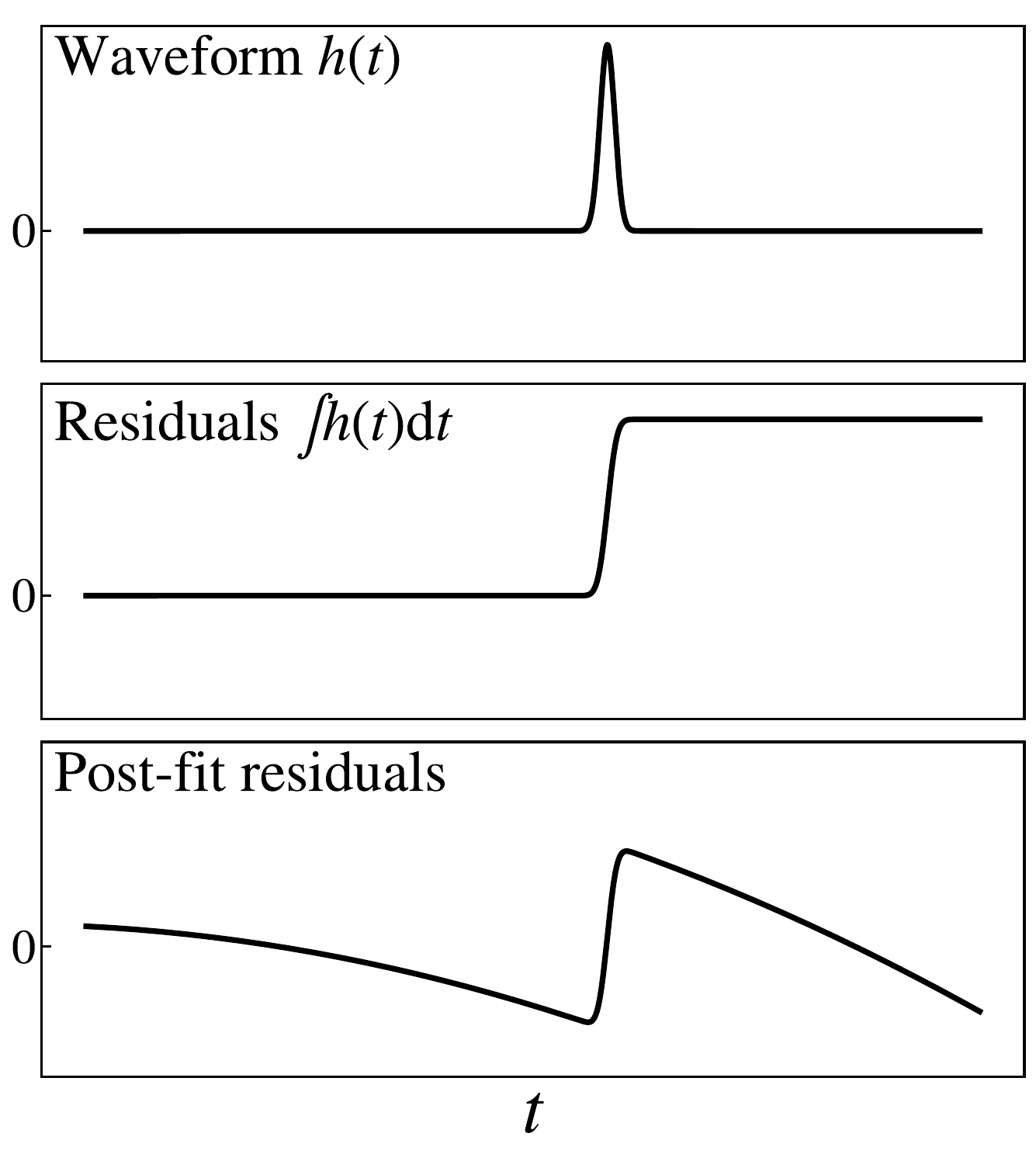}
    }
    \subfloat[]{%
        \includegraphics[width=0.25\textwidth]{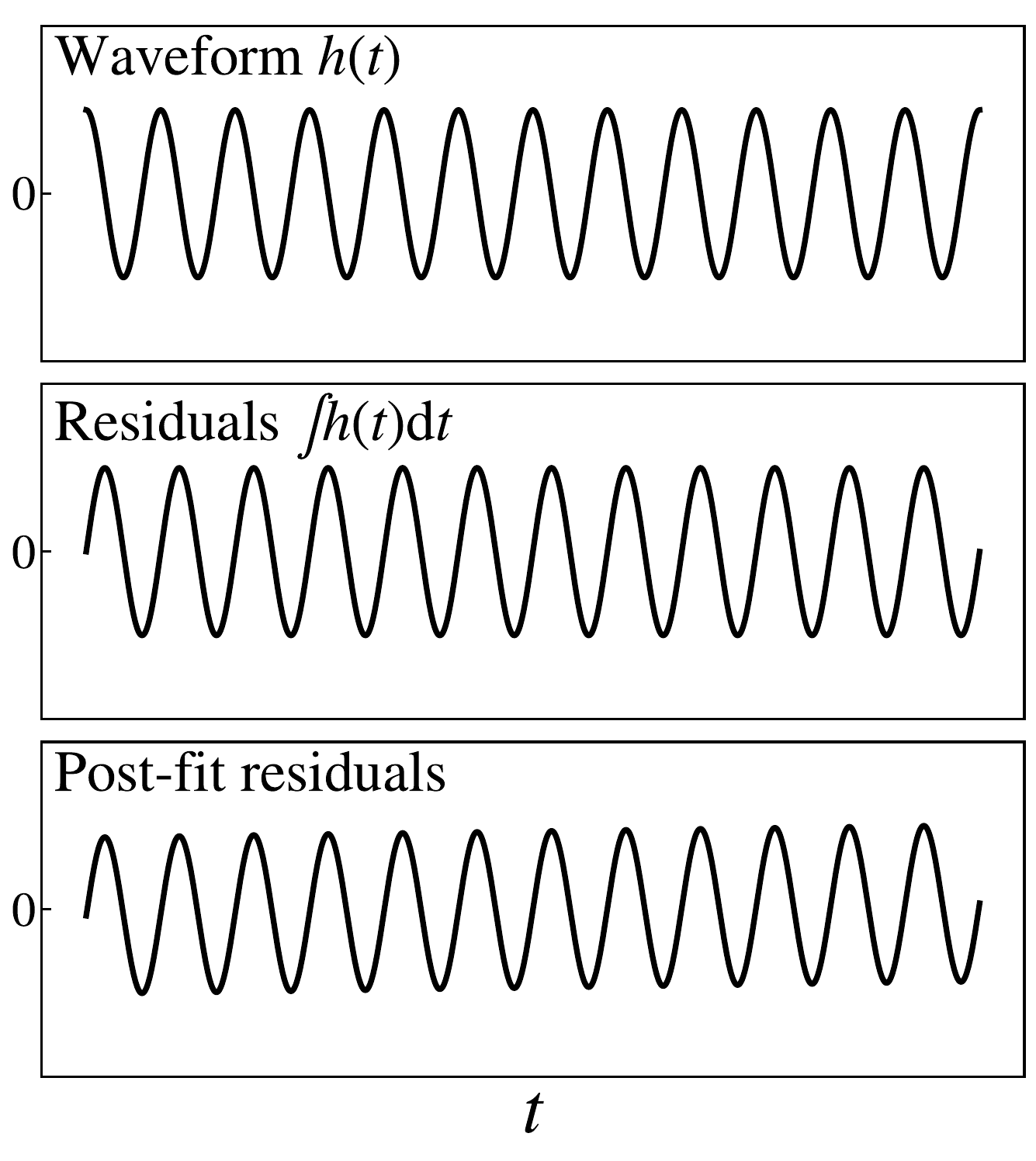}
    }
    \caption{%
        Illustration of four sine-Gaussian waveforms, the induced residuals, and the residuals after the quadratic trend is removed (post-fit). (a) A wave-packet.
        (b) A signal resembling the parabolic encounter of two SMBHs. (c) A short burst. (d) A continuous sinusoidal signal.
    }
    \label{fig:four_panel}
\end{figure*}

If the Gaussian width $\sigma_{\rm sg}$ is well within the observation span and sinusoidal oscillations occur within the envelope, the signal is a Gaussian wave packet. Both the induced and post-fit residuals retain a wave packet. Fig.~\ref{fig:four_panel}(a) shows the waveform, the induced residuals, and the post-fit residuals for a signal arriving near the midpoint of the dataset. In this case, the removal of the timing model's quadratic components has little effect on the signal shape. In Sections \ref{sine} and \ref{sine-Gaussian}, we use the sine-Gaussian wave packet to model the pulsar transient noise.

\subsection{Encounter of SMBHs}
If the Gaussian envelope contains only a single oscillation, the waveform resembles that produced during the close encounter of two SMBHs in a highly eccentric orbit. The corresponding induced and post-fit residuals are shown in Fig. \ref{fig:four_panel}(b). Here, we consider a signal arriving later than the midpoint of the observation time. Removal of the quadratic components introduces asymmetry in the post-fit residuals.

\subsection{Short burst}
Setting the phase to $\phi_{\rm sg}=\pi/2$ (so that the sine becomes a cosine) and choosing a Gaussian width much smaller than the oscillation period lead to a sharply peaked waveform. The induced residuals take the form of an approximate step function, as shown in Fig.~\ref{fig:four_panel}(c). This case is analogous to a short burst from, e.g., a cosmic string cusp or any high-frequency transient whose waveform cannot be fully resolved by the discrete TOAs. As in the previous example, we consider a signal arriving later than the midpoint of the observation span.

\subsection{Continuous waves}
If the Gaussian width is much larger than the observational span, the waveform reduces to an approximately monochromatic sinusoid. Such a signal corresponds to the Earth-term–only, non-evolving continuous waves from a circular SMBHB. The waveform, induced residuals, and post-fit residuals all have nearly identical sinusoidal forms, as shown in Fig.~\ref{fig:four_panel}(d).

These examples demonstrate that a variety of physically interesting GW signals can be represented as smooth functions in the time domain and are therefore naturally captured by a Fourier expansion with a modest number of components. This observation underlies the waveform-agnostic approach developed in the following sections.

\end{document}